\documentclass[a4paper,12pt]{article}

\usepackage{cite}
\usepackage{ulem}
\usepackage{amsmath,amssymb}
\usepackage{graphicx}
\usepackage{color}

\DeclareMathOperator{\sgn}{sgn}

\DeclareMathOperator{\U}{\text{U}}

\usepackage{color}
\usepackage{geometry}
\numberwithin{equation}{section}
\allowdisplaybreaks

\begin{document}
\newcommand{\aff}[1]{${}^{#1}$}
\renewcommand{\thefootnote}{\fnsymbol{footnote}}

\begin{titlepage}
\begin{flushright}
{\footnotesize YITP-18-99}
\end{flushright}
\begin{center}
{\Large\bf
Aspects of Massive Gauge Theories 
on Three Sphere \\ \vspace{1mm}  in Infinite Mass Limit
}
\\
\bigskip\bigskip
{\large Kazuma Shimizu\footnote{
\tt kazuma.shimizu(at)yukawa.kyoto-u.ac.jp
}\aff{1}
}\\
\bigskip\bigskip
\aff{1}: {\small
\it Yukawa Institute for Theoretical Physics, Kyoto University, Kyoto 606-8502, Japan
}
\end{center}

\begin{abstract}
{
We study the $S^3$ partition function of three-dimensional supersymmetric $\mathcal{N}=4$ U($N$) SQCD with massive matter multiplets in the infinite mass limit with the so-called Coulomb branch localization. We show that in the infinite mass limit a specific point of the Coulomb branch is selected and contributes dominantly to the partition function. Therefore, we can argue whether each multiplet included in the theory is effectively massless in this limit, even on $S^3$, and conclude that the partition function becomes that of the effective theory on the specific point of the Coulomb branch in the infinite mass limit. In order to investigate which point of the Coulomb branch is dominant, we use the saddle point approximation in the large $N$ limit because the solution of the saddle point equation can be regarded as a specific point of the Coulomb branch. Then, we calculate the partition functions for small rank $N$ and confirm that their behaviors in the infinite mass limit are consistent with the conjecture from the results in the large $N$ limit. Our result suggests that the partition function in the infinite mass limit corresponds to that of an interacting superconformal field theory.

 }
\end{abstract}

\bigskip\bigskip\bigskip

\end{titlepage}

\renewcommand{\thefootnote}{\dag\arabic{footnote}}
\setcounter{footnote}{0}

\tableofcontents
\newpage

\section{Introduction}

In three dimensions, the Yang-Mills coupling has positive mass dimension. This means that three-dimensional Yang-Mills theories are super-renormalizable. The Yang-Mills term is irrelevant and cannot contribute to the infrared physics independently of the gauge group and the matter content. It might be expected that 3d gauge theories flow to the non-trivial infrared fixed point, which depends on the matter content. In fact, U($N$) QCD with $N_{f}\geq N_{\text{cri}}$ massless flavors, where $N_{\text{cri}}$ is some critical value, might flow to an interacting IR fixed point while with $N_{f} < N_{\text{cri}}$ massless flavors the theory is expected to flow to a gapped phase in the IR \cite{Pi,ANW1,ANW2}. The numbers of the flavors plays an important role in determining the IR structure of a 3d gauge theory. However, it is generally difficult to determine the non-perturbative properties of such a theory.

Three-dimensional supersymmetric gauge theories have several interesting features that four-dimensional supersymmetric gauge theories do not share. In particular, we are interested in the fact that there are real parameters, namely the real mass and Fayet-Iliopolous (FI)  parameters. These are not given by the background chiral superfields. Thus, the dynamics triggered by the deformation of real parameters are not restricted by holomorphy. This means that the non-trivial phase transitions can occur. When we give the matter fields infinite mass, the massive matter fields decouple from the theory and decoupling of the flavors changes the IR physics. Then, an interesting phase transition occurs.

Supersymmetric gauge theories are known to have exactly calculable quantities such as the partition function on a compact manifold $\mathcal{M}$ using localization methods in three dimensions \cite{KWY,KWY2,Ja,HHL}. In this paper, we focus on the round three-sphere partition function, which is given by a matrix-type finite-dimensional integral. These localization methods admit the deformation of the real mass and FI terms by weakly gauging a global symmetry and giving the background field that couples with the current of global symmetry an expectation value. Then, we can approach the non-trivial dynamics triggered using these real mass parameter with the localization methods. In \cite{BR,RST,AZ,AR,NST1,NST2,RT,AD,HNST,ST,ARo,ST1}, the phase structure of mass-deformed gauge theories on $S^{3}$ is investigated.

In this study, we focus on $\mathcal{N}=4$ U($N$) SQCD with $N_{f}$ pairs of chiral multiplets in the fundamental and anti-fundamental representations of U($N$). These theories are classified in \cite{GW} by their low energy properties. The authors define three types of the theories: ``good", ``ugly" and ``bad" theories.
A 3d gauge theory is a good theory if all the monopole operators obey the unitarity bound. In this case, the R-symmetry in the IR is the same as that in the UV. An $\mathcal{N}=4$ U($N$) SQCD is a good theory when $N_{f}\geq 2N$. A gauge theory is called ``ugly" if the monopole operators satisfy the unitarity bound, but somel monopole operators saturate it. This type of theory is likely to flow to an interacting superconformal field theory (SCFT) with R-symmetry visible in the UV and a decoupled free sector consisting of the monopole operators that saturate the unitarity bound. An $\mathcal{N}=4$ U($N$) SQCD is an ugly theory when $N_{f}=2N-1$. In a bad theory, there are monopole operators with zero or negative R-charge corresponding to the R-symmetry manifest in the UV. Because the monopole operators violate the unitarity bound of the UV R-symmetry, a bad theory flows to a fixed point, whose R-symmetry is not manifest in the UV. An $\mathcal{N}=4$ U($N$) SQCD becomes a bad theory when $N\leq N_{f}\leq 2N-2$ \footnote{Recent progress on``bad" theories in terms of the geometry of the moduli space of vacua is described in \cite{AC,DK,AC2}.}. It is known that the question of whether the $S^{3}$ partition function diverges is related to the criterion of ``bad" theories. The partition function of a ``bad" theory is divergent \cite{KWY2}. This might be because the localization methods use the R-symmetry that is manifest in the UV to define the gauge theory on a compact manifold. Thus, we should treat the number of the flavors carefully.

Our aim in this work is to study the $S^{3}$ partition function of real mass-deformed theories in the infinite mass limit\footnote{
The infinite mass limit of the matrix model of 3d gauge theories is also considered in \cite{WY,Ni,Yaakov,ARSW,Ama} in the context of finding new examples of Seiberg-like dualities \cite{Aha,GK}.
}. For example, we consider the situation in which we give real masses to enough matter multiplets of a  ``good" theory for it to become a``bad" theory after the massive matter fields decouple. It could naively be thought that the massive matter multiplets will decouple from the theory in this limit. However, a matrix model of a ``bad" theory is not well defined\footnote{
The magnetic theory of a ``bad" theory in terms of the Seiberg-like duality is considered as a good theory \cite{Yaakov}.}. It is interesting to investigate what happens to this matrix model in the infinite mass limit. Hence, our interest is to determine which hypermultiplets become effectively massless or massive in the infinite mass limit on the three-sphere. When a theory is defined on the flat space, we must choose a vacuum in order to determine the decoupling of matter fields. However, there are no vacuum choices for the theories on the three-sphere. In particular, we calculate the sphere partition function with the help of so-called Coulomb branch localization and this is given by the integral over the classical Coulomb branch parameters. Namely, the three-sphere partition function is represented by the integrals of a portion of the vacua in terms of the theory on flat space. Thus, it is not simple to argue whether or not the massive multiplets will be decoupled when we take the infinite mass limit.

\begin{figure}[t!]
\label{CBgra}
 \begin{center}
 \includegraphics[width=8cm]{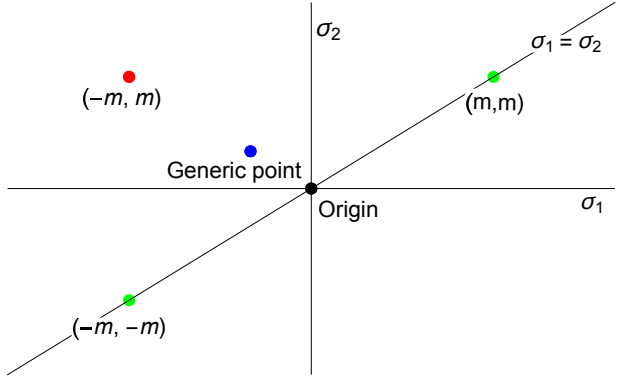}
 \label{zu}
 \end{center}
 \caption{
This figure schematically shows the real parts of the two classical Coulomb branch parameters of U(2) $\mathcal{N}=4$ SQCD with $\frac{N_f}{2}$ pairs of hypermultiplets with real mass $\pm m$. There are some special points where new massless degrees of freedom appear or the gauge symmetry is enhanced to U(2). Here, we assume that $\sigma_2\geq \sigma_1$ due to the Weyl symmetry of U(2).}
 \end{figure}

 For example, we consider U(2) $\mathcal{N}=4$ SQCD with $\frac{N_f}{2}$ pairs of hypermultiplets with a real mass $\pm m$. Figure \ref{CBgra} shows the real parts of the two classical Coulomb branch parameters, where there are some special points. When we fix a generic point of the Coulomb branch (blue dot), the effective theory is U(1)$\times$ U(1) with massive matter fields and W-bosons while on a specific point, such as green or red points, the effective theory has $\frac{N_f}{2}$ or $N_f$ massless hypermultiplets, respectively. The origin (black dot) is also special in the sense that the gauge symmetry is enhanced to U(2). It is non-trivial to determine which points dominantly contribute to the three-sphere partition function in the infinite mass limit, because all the points of the Coulomb branch can contribute to it, including generic points and the special one mentioned above.

To investigate this, we focus on the solution of the saddle point equation because the solution corresponds to a classical Coulomb branch point and in the large $N$ limit it gives a dominant contribution to the sphere partition function. Hence, in the large $N$ limit, we can investigate the decoupling of the massive matter fields as well as which theory will appear as an effective theory on the point of the Coulomb branch which corresponds to the solution. Then, we deduce the effective theory in the infinite mass limit because the solution of the saddle point equation of the effective theory coincides with that of the saddle point equation of the original massive theory in the infinite mass limit.   

Investigating the solution of the saddle point equation is simply a methods to determine which point of the  Coulomb branch gives the dominant contribution to the partition function in the infinite mass limit. Even when we do not take the large $N$ limit, it is expected that there exists a dominant point of the Coulomb branch and the matrix model becomes a specific effective theory in the infinite mass limit. This is because the mass infinite limit also corresponds to the decompactified limit $(r_{S^3} \to \infty)$\footnote{The mass $m$ must appear as the combination $mr_{S^3}$ in the partition function. Therefore we cannot distinguish between the infinite mass limit and the decompactified limit. In our convention we take $r_{S^{3}}$ to 1.}
 and thus the point of the Coulomb branch should be chosen in this limit. We verify this in the matrix models for small $N$ and confirm that the effective theory is the same as that which we deduced from the calculations in the large $N$ limit. We conclude that this vacuum selection does not require the large $N$ limit, rather than just the infinite mass limit.

 In this study, we focus on the following two types of the mass-deformed $\mathcal{N}=4 $ U($N$) SQCD: (i) with only massive matter fields, and (ii) with massive and massless matter fields. The theory (i) is simple and suitable to investigate the cases in which the mass deformation leads to a `bad' theory after decoupling of the matter fields. The theory (ii) is also simple and suitable for investigating whether the massive matter fields simply decouple when the mass deformation leads a `good' theory to a `good' theory and how the gauge group is spontaneously broken as the number of the massless fields increases. In addition, we can obtain some insight into the results even in case of general mass deformations from those in case of these deformations.

The rest of this paper is organized as follows: In Section 2, we review localization methods and introduce the building blocks of matrix models. In Section 3, we solve the saddle point equation of $\mathcal{N}=4$ SQCD with massless or massive matter fields and investigate the theory that appears in the infinite mass limit. In Section 4, we calculate the partition function of finite rank SQCDs and evaluate the leading part in the infinite mass limit. In Section 5 we present a conclusion and discussion. In Appendix \ref{res}, we introduce the techniques of the resolvent methods utilized in this paper to solve the saddle point equation in the large $N$ limit. In Appendix \ref{mix}, we introduce mixed Chern-Simons terms which must appear in the infinite mass limit as one-loop effects. We attempt to interpret what happens in the infinite mass limit in terms of these mixed Chern-Simons terms. In Appendix \ref{conv}, we discuss the convergence bound of the matrix model and reconsider the matrix model of the effective theory in the infinite mass limit from the viewpoint of its convergence bound. In Appendix \ref{ABJM}, we introduce an example that becomes ABJM theory in the infinite mass limit while it is just an SQCD when $m=0$.

\section{Localization and matrix model}

In this paper, we investigate the round three-sphere partition function of three dimensional supersymmetric gauge theories. This is given by a finite dimensional integral rather than the path integral by employing the localization technique \cite{KWY,KWY2,Ja,HHL}. To utilize this localization technique, we define a gauge theory on $S^{3}$ with preserving supersymmetry and deform the action on $S^{3}$ by a Q-exact term, where Q is a generator of the supersymmetry. The partition function of the deformed action is independent of the deformation parameter.
Thus, we take the parameter to infinity and the path integral reduces to a finite-dimensional matrix integral because the path integral is determined by the finite-dimensional saddle point configuration in the field configuration space. Because the saddle point approximation is one-loop exact, the action of the matrix model is written in terms of classical and one-loop parts as
\begin{align}
Z=\frac{1}{N!}\int \left(\prod_{i=1}^{\text{Rank}(G)}d\sigma_{i}\right)|J| Z_{\text{Classical}}(\sigma)Z^{\text{vec}}_{\text{1-loop}}(\sigma)Z^{\text{mat}}_{\text{1-loop}}(\sigma),
\end{align}
where $|J|$ is the usual Vandermonde determinant and $Z^{\text{vec}}_{\text{1-loop}}$ and $Z^{\text{mat}}_{\text{1-loop}}$
are one-loop contributions  from the vector multiplets and matter multiplets, respectively. The variable $\sigma_{i}$ of the integral corresponds to an eigenvalue of the scalar fields of an $\mathcal{N}=2$ vector multiplet.

\subsection{Vector multiplet}
In this paper, we use the Coulomb branch localization mentioned above. We only consider U($N$) gauge theories in this paper. The Yang-Mills term cannot contribute to the partition function since the Yang-Mills term is Q-exact. On the other hand, the Chern-Simons term can contribute to this as a classical contribution, but we do not consider this situation here. The one-loop contribution of the vector multiplets is given by 
\begin{align}
Z^{\text{vec}}_{\text{1-loop}}(\sigma)=\prod_{i<j}^{N}\frac{4\sinh^2\pi (\sigma_{i}-\sigma_{j})}{\pi^2(\sigma_{i}-\sigma_{j})^2},
\end{align}
where the denominator cancels against the Vandermonde determinant, which appears when we choose the diagonal gauge of $\sigma$. When there is a U(1) part of the gauge group, the FI term can be introduced, which contributes to the partition function as a classical term
\begin{align}
e^{2\pi i \zeta\sum_{i=1}^{N}\sigma_{i}}.
\end{align}

\subsection{Matter multiplet}
Next, we consider the contributions of chiral multiplets. The chiral multiplets can only contribute to the partition function through one-loop parts because their kinetic and superpotential terms are Q-exact. The one-loop contributions of chiral multiplets are determined by the representations of both the gauge group and the flavor symmetry. By weakly gauging a flavor symmetry, we can couple its current with a background vector multiplets in a supersymmetric manner. Thus, we can give the corresponding scalar $\sigma_{\text{b}}$ an expectation value and regard it as a real mass for the chiral multiplets. Moreover, we can give the chiral multiplets an R-charge \cite{Ja,HHL}. However, we do not consider such a deformation in this work and we consider the chiral multiplets to have canonical dimension $\frac{1}{2}$ \footnote{ For an $\mathcal{N}=4$ vector multiplet, there exists an adjoint chiral multiplet in terms of the $\mathcal{N}=2$ language. Then, it appears that we must consider the one-loop contributions of this. However, because its canonical R-charge is 1, the adjoint chiral multiplet does not contribute to the partition function without an axial mass parameter \cite{KWY2,HHL}.}.

The one-loop contribution of the chiral multiplets in the representation $R$ of U($N$) is given by
\begin{align}
\prod_{\rho}e^{\ell\left(\frac{1}{2}+i\rho(\sigma)\right)},
\end{align}
where $\rho$ is a weight vector of the representation $R$. In \cite{Ja}, the function $\ell (z)$ is defined as  
\begin{align}
\ell(z)=-z\log\left(1-e^{2\pi i z}\right)+\frac{i}{2}\left(\pi z+\frac{1}{\pi}\text{Li}_{2}(e^{2\pi iz })\right)-\frac{i\pi }{12}.
\end{align}
A notable property we will often exploit is that 
\begin{align}
e^{\ell(\frac{1}{2}+ix)}e^{\ell(\frac{1}{2}-ix)}=\frac{1}{2\cosh\pi x}.
\end{align}

In this paper, we focus on SQCDs, which are super Yang-Mills theories with $ N_{f}$ pairs of chiral multiplets in the fundamental and anti-fundamental representations of U($N$). We consider following two mass deformations: In case (i), we give a real mass $m$ to $\frac{N_{f}}{2}$ flavors while we give a real mass $-m$ to the remaining $\frac{N_{f}}{2}$ flavors\footnote{We assume that $\frac{N_{f}}{2}$ is an integer. }.
This breaks flavor symmetry SU($N_{f}$) down to SU($\frac{N_{f}}{2}$)$\times$SU($\frac{N_{f}}{2}$). The total one-loop contribution is given by 
\begin{align}
Z^{\text{mat}}_{\text{1-loop}}(\sigma)=&\prod_{i=1}^{N}e^{\frac{N_{f}}{2}\left(\ell\left(\frac{1}{2}+i(\sigma_{i}+m)\right)+\ell\left(\frac{1}{2}+i(\sigma_{i}-m)\right)+\ell\left(\frac{1}{2}+i(-\sigma_{i}+m)\right)+\ell\left(\frac{1}{2}+i(-\sigma_{i}-m)\right)\right)}\nonumber \\
=&\prod_{i=1}^{N}\frac{1}{2\left(\cosh\pi\left(\sigma_{i}+m\right)2\cosh\pi\left(\sigma_{i}-m\right)\right)^{\frac{N_{f}}{2}}}.
\end{align}
In case (ii), we give $\frac{N_f}{3}$ flavors a real mass $m$ while we give other $\frac{N_f}{3}$ flavors a real mass $-m$. Then, we keep the remaining $\frac{N_{f}}{3}$ flavors massless. This real mass assignment breaks each of the SU($N_{f}$) global symmetries of the matter fields down to SU($\frac{N_f}{3}$)$\times$SU($\frac{N_f}{3}$)$\times$SU($\frac{N_f}{3}$)\footnote{We assume that $\frac{N_{f}}{3}$ is an integer.}. The total one-loop contribution of the chiral multiplets is given by 
\begin{align}
Z^{\text{mat}}_{\text{1-loop}}(\sigma)=&\prod_{i=1}^{N}e^{\frac{N_{f}}{3}\left(\ell\left(\frac{1}{2}+i(\sigma_{i}+m)\right)+\ell\left(\frac{1}{2}+i(\sigma_{i}-m)\right)+\ell\left(\frac{1}{2}+i(-\sigma_{i}+m)\right)+\ell\left(\frac{1}{2}+i(-\sigma_{i}-m)\right)+\ell\left(\frac{1}{2}+i\sigma_{i}\right)+\ell\left(\frac{1}{2}-i\sigma_{i}\right)\right)}\nonumber \\
=&\prod_{i=1}^{N}\frac{1}{\left(2\cosh\pi\left(\sigma_{i}+m\right)2\cosh\pi\left(\sigma_{i}-m\right)2\cosh\pi \sigma_{i}\right)^{\frac{N_{f}}{3}}}.
\end{align}

\section{Large $N$ solution and Coulomb branch point}
\label{sec3}

\subsection{SQCD with massless hypermultiplets}
In this subsection, we solve the saddle point equation of U($N$) SQCD with massless hypermultiplets for later use. The solution is given as an eigenvalue density function $\rho(x)$, which determines the large $N$ behavior of the theory. The partition function is written as 
\begin{align}
Z=\frac{1}{N!}\int \prod_{i=1}^{N}dx_{i}\frac{\prod_{i<j}4\sinh^2\left(\pi (x_{i}-x_{j})\right)}{\prod_{i}\left(2\cosh(\pi x_{i})\right)^{N_{f}}}.
\end{align}
It is generally difficult to calculate this partition function exactly except for small $N$. Fortunately, the leading part in the large $N$ limit can be evaluated by the saddle point approximation. The saddle point equation for this theory is given by 
\begin{align}
0=N_f\tanh(\pi x_{i})-2\sum_{j \neq i}\coth\pi(x_{i}-x_{j}).
\end{align}
We assume that the eigenvalues become dense in the large $N$ limit and we take the continuous limit as follows:
\begin{align}
 \frac{i}{N}\rightarrow s\in[0,1],\quad x_{i} \rightarrow x(s),\quad \frac{1}{N}\sum_{i=1}^{N}\rightarrow \int ds. 
\end{align}
  The leading part of this saddle point equation is rewritten as a singular integral equation\footnote{
We denote a principal value integral as 
\begin{align}
\text{P}\int dx.\nonumber
\end{align} 
  }
\begin{align}
\label{resolv}
0=\xi \tanh\pi(x)-2\left(\text{P}\int dy \rho(y)\coth\pi(x-y)\right),
\end{align}
where we also took $N_{f}$ to be infinite with $\xi\equiv\frac{N_f}{N}$ finite and introduced the density function $\rho(x)$ defined as 
\begin{align}
\frac{ds}{dx}\equiv \rho(x).
\end{align}
This means that we regard the values of the eigenvalues denoted by $x$ as constituting the fundamental variables. The density function $\rho(x)$ counts the number of the eigenvalues which exist between $x$ and $x+dx$ and satisfies the following normalization condition which depends on how we take the continuous limit:
\begin{align}
\int_{I}dx \rho(x)=1.
\end{align}
In order to solve the equation \eqref{resolv} and obtain the density function $\rho(x)$, we employ the resolvent methods. We give a brief summary of resolvent methods in Appendix \ref{res}. We take $e^{2\pi x}\equiv X$ and $e^{2\pi y}\equiv Y$ and define the resolvent $\omega(Z)$ and the potential $V^{\prime}(x)$ as
\begin{align}
\omega(X)\equiv &2\int_{I} dy \rho(y) \frac{e^{\pi (x-y)}+e^{-\pi (x-y)}}{e^{\pi (x-y)}-e^{-\pi (x-y)}}=2\int_{I} dy\rho(y)\frac{X+Y}{X-Y}=2\bigg(1+\int_{\mathcal{C}} \frac{dY}{\pi}\frac{\rho(y)}{X-Y}\bigg), \\
V^{\prime}(x) \equiv& \frac{X-1}{X+1}\xi,
\end{align}
where $I$ and $\mathcal{C}$ represent the  intervals $[x_{\text{min}},x_{\text{max}}]$ and $[b,a]$ respectively. The resolvent is determined from the analyticity and the one-cut solution of the resolvent is given by \eqref{omega1} as
\begin{align}
\label{formula1}
\omega(X)=\xi\left(\frac{X-1}{X+1}-\frac{2\sqrt{(X-a)}\sqrt{(X-\frac{1}{a})}}{(X+1)\sqrt{(1+a)(1+\frac{1}{a})}}\right)=\omega_{0}\left(X;1;a,\frac{1}{a}\right),
\end{align}
where $b=\frac{1}{a}$ because of the symmetry of the saddle point equation. We should carefully consider the branch of the square root. For later convenience, we introduce the following function:
\begin{align}
\omega_{0}(X;A;a,b)=\xi\left(\frac{X-A}{X+A}-\frac{2A\sqrt{(X-a)}\sqrt{(X-b)}}{(X+A)\sqrt{(1+a)(1+b)}}\right).
\end{align}
The density function $\rho(x)$ defined on $[\frac{1}{a},a]$ is given by \eqref{analy2} as
\begin{align}
\label{rho1}
\rho(x)
=& \frac{\xi}{(X+1)}\sqrt{\frac{(X-\frac{1}{a})(a-X)}{(1+a)(1+\frac{1}{a})}}.
\end{align}
The end of the cut $a$ is determined by the asymptotic behavior of the resolvent $\omega(X)$ from \eqref{formula1}. The asymptotic behavior in $X\rightarrow\ 0$ is determined by the following equation:
\begin{align}
\label{xi}
\frac{-2}{\xi}=-1+\frac{2}{\sqrt{(1+a)(1+\frac{1}{a})}}.
\end{align}
The solution is given by
\begin{align}
\label{length1}
a=\frac{\xi^2+4\xi-4+4\sqrt{(\xi-1)\xi^2}}{(\xi-2)^2},
\end{align}
where this solution only exists when $\xi\geq 2$ because the right-hand side of \eqref{xi} is always greater than $-1$ as a function of $a$.

Here, we argue on the relation between this large $N$ solution and a point of the classical Coulomb branch. The equation \eqref{length1} implies that when we take $r_{S^{3}}$ to infinity, $x_{\text{min}}$ and $x_{\text{max}}$ become 0 because the radius is recovered as $x\rightarrow xr_{S^{3}}$ and $a=e^{2\pi r_{S^{3}}x_{\text{max}}}$. Thus, the saddle point solution becomes condensed to the origin. Taking the radius to infinity corresponds to considering the theory on a flat space. Therefore, this solution corresponds to a point of the Coulomb branch of the theory on a flat space, which is the origin of the classical Coulomb branch. The origin of the Coulomb branch is the most singular point in the sense that on this point, all the massive W-bosons become massless. On this point, the theory at the deep IR of the RG flow expected to be an interacting superconformal field theory. It is expected that the sphere partition function of SQCD with massless hypermultiplets always represents that of a non-trivial SCFT.

The solution exists when $\xi \geq 2$. This reflects the bound of the convergence of the matrix model.
In this study, we will add real mass to matter fields while preserving the special flavor symmetry. Even in that case, this bound always appears in our analysis.

\subsubsection{Adding an FI term}
\label{FI}
Here, we consider U($N$) gauge theories with $N_{f}$ massless hypermultiplets and an FI term. In particular in this section, we consider imaginary FI terms. This is in preparation for the latter part of this paper, where such terms appear as one-loop effects when we take the infinite mass limit, namely in the form of certain mixed Chern-Simons terms. The density function is almost the same as that in the previous section. However, an FI term breaks the symmetry of the saddle point equation under the simultaneous change of the sign of all eigenvalues $x_{i}\rightarrow -x_{i}$.

For this theory, the matrix model is written as
\begin{align}
\label{FImat}
Z=\frac{1}{N!}\int \prod_{i=1}^{N}dx_{i}\frac{e^{\pi \zeta \sum_{i}x_{i}}\prod_{i<j}4\sinh^2\left(\pi (x_{i}-x_{j})\right)}{\prod_{i}\bigg(2\cosh\pi (x_{i})\bigg)^{N_{f}}},
\end{align}
where $\zeta\in \mathbb{R}$ is an imaginary FI parameter in the sense that ordinary FI terms are considered as $e^{i\zeta \pi\sum_{i}x_{i}}$. This FI term can be considered as the R-charge of the monopole operator since the real part of the monopole operator is $e^{-2\pi \Delta_{m} \sigma_{i}}$, where $\Delta_{m}$ is the R-charge of the monopole operator \cite{JKPS,LY}. The saddle point equation in the continuous limit is 
\begin{align}
0=\eta+\xi\tanh\pi x_{i}-2\left(\text{P}\int_{I} dy\rho(y)\coth\pi(x-y)\right),
\end{align}
where we also take $N_{f}$ and $\zeta$ to be infinite while keeping 
\begin{align}
\xi\equiv \frac{N_{f}}{N},\quad \eta\equiv \frac{\zeta}{N},
\end{align}
finite in order to solve the saddle point equation. We can solve this saddle point equation in the large $N$ limit using resolvent methods. We define the resolvent $\omega(X)$ and potential $V^{\prime}(x)$ for this theory as
\begin{align}
\omega(X)\equiv &2\int dy\rho(y)\frac{X+Y}{X-Y}=2\bigg(1+\int \frac{dY}{\pi}\frac{\rho(y)}{X-Y}\bigg), \\
V^{\prime}(x) \equiv& \eta+\frac{X-1}{X+1}\xi.
\end{align}
The resolvent $\omega(Z)$ is obtained through the same calculation that appears in the previous section because an FI term does not change the singular structure of the resolvent:
\begin{align}
\omega(X)=\eta+\omega_{0}\left(X;1;a,b\right).
\end{align}
The density function is given by the equation \eqref{analy2} as
\begin{align}
\label{FIdensity}
\rho(x)
=&\xi\bigg[\frac{\sqrt{(a-X)(X-b)}}{(X+1)\sqrt{(1+a)(1+b)}}\bigg],
\end{align}
where $a$ and $b$ are determined from the equation describing the asymptotic behavior of $\omega(X)$ at $X=0$ and $\infty$:
\begin{align}
\label{eta1}
\frac{\eta}{\xi}=&\frac{1-\sqrt{ab}}{\sqrt{(1+b)(1+a)}},\\
\label{eta2}
1-\frac{2}{\xi}=&\frac{1+\sqrt{ab}}{\sqrt{(1+b)(1+a)}}.
\end{align}
Because an FI term breaks the $\mathbb{Z}_{2}$ symmetry under which $x_{i} \rightarrow -x_{i}$ in the saddle point equation, $a$ and $b$ do not satisfy the condition $ab=1$. The solutions of \eqref{eta1} and \eqref{eta2} are given by
\begin{align}
\label{soleta1}
a=\frac{-4 +4\xi+\xi^2-\eta^2+4 \sqrt{(\xi-1)\left(
\xi^2-\eta^2\right)}}{(-2 +\xi+\eta)^2},\quad 
b=\frac{-4 +4\xi+\xi^2-\eta^2-4 \sqrt{(\xi-1)\left(
\xi^2-\eta^2\right)}}{(-2 +\xi+\eta)^2} .
\end{align} 
From \eqref{eta1} and \eqref{eta2}, we find that the solution only exists when 
\begin{align}
\label{exist2}
\xi\geq 2+|\eta|.
\end{align} 
This condition is equivalent to the condition that the matrix model converges in the large $N$ limit. In Appendix \ref{conv}, we will discuss the convergence bound of the matrix model of SQCDs.

\subsection{SQCD with massive hypermultiplets}
\label{32}
In this subsection, we consider U($N$) SQCD with $N_{f}$ pairs of chiral multiplets with real mass by weakly gauging its flavor symmetry and coupling its current to $\mathcal{N}=2$ vector multiplets as background fields such that the matrix model is given by
\begin{align}
\label{matrixqcd2}
Z=\frac{1}{N!}\int \prod_{i=1}^{N}dx_{i}\frac{\prod_{i<j}4\sinh^2\left(\pi (x_{i}-x_{j})\right)}{\prod_{i}\bigg(2\cosh\pi (x_{i}+m)2\cosh\pi (x_{i}-m)\bigg)^{\frac{N_{f}}{2}}}.
\end{align}
When $m=0$, this matrix model becomes that of U($N$) with $N_{f}$ massless fundamental hypermultiplets. When we take the infinite mass limit, if the massive matter multiplets decouple, then, the matrix model is not well defined. Therefore, we investigate what happens to this matrix model in the infinite mass limit.

 The saddle point equation is written as 
\begin{align}
\label{sadqcd2}
2\sum_{i}\coth\pi(x_{i}-x_{j})=\frac{N_{f}}{2}\bigg(\tanh\pi(x_{i}+m)+\tanh\pi(x_{i}-m)\bigg),
\end{align}
and in the continuous limit this becomes 
\begin{align}
4\left(\text{P}\int dy\rho(y)\coth\pi(x-y)\right)=\xi\bigg(\tanh\pi(x+m)+\tanh\pi(x-m)\bigg),
\end{align}
Next, we define the resolvent $\omega(X)$ and potential $V^{\prime}(x)$ as
\begin{align}
\omega(X)=&4 \int dy \rho(y)\frac{X+Y}{X-Y}=4\bigg(1+\int \frac{dY}{\pi}\frac{\rho(y)}{X-Y}\bigg),\\
V^{\prime}(x)=&\xi\left(\frac{X-M^{-1}}{X+M^{-1}}+\frac{X-M}{X+M}\right),
\end{align} 
where $M=e^{2\pi m}$. The resolvent is determined by its analytic properties \eqref{omega1} as 
\begin{align}
\label{formula2}
\omega(X)=&\omega_{0}\left(X;M;a,b\right)+\omega_{0}\left(X;M^{-1};a,b\right).
\end{align}
The density function is given by \eqref{analy2} as
\begin{align}
\label{densitysqcd3}
\rho(x)
=&\frac{\xi}{2}\bigg[\frac{M\sqrt{(a-X)(X-b)}}{(X+M)\sqrt{(M+a)(M+b)}}+\frac{M^{-1}\sqrt{(a-X)(X-b)}}{(X+M^{-1})\sqrt{(M^{-1}+a)(M^{-1}+b)}}\bigg].
\end{align}
The constants $a$ and $b$ are detemined by the symmetry and asymptotic behavior when $Z=0$:
\begin{align}
-4=2\xi\left(-1+\frac{1}{\sqrt{(M+a)(M+\frac{1}{a})}}+\frac{1}{\sqrt{(M^{-1}+a)(M^{-1}+\frac{1}{a})}}\right).
\end{align}
This equation immediately implies that $a$ exists when $\xi\geq2$. We conclude that this type of mass deformation does not affect the bound of the existence of the solution. Here, $a$ is given by
\begin{align}
\label{interbal1}
a=\frac{2(\xi-1)(M^2+1)+M\xi^2+2(M+1)\sqrt{(\xi-1)\left(\xi-1+M^2(\xi-1)+M(\xi^2-2\xi+2)\right)}}{M (\xi-2)^2}.
\end{align}

\begin{figure}[ht!]
 \begin{tabular}{cc}
 \begin{minipage}{0.5\hsize}
 \begin{center}
 \includegraphics[width=8cm]{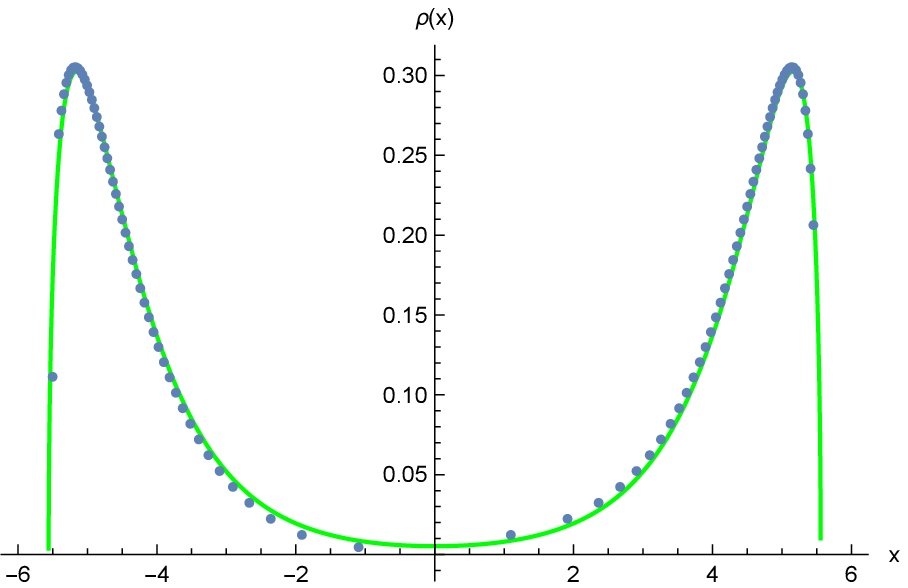}
 \label{nccomp1}
 \end{center}
 \end{minipage}
 \begin{minipage}{0.5\hsize}
 \begin{center}
 \includegraphics[width=8cm]{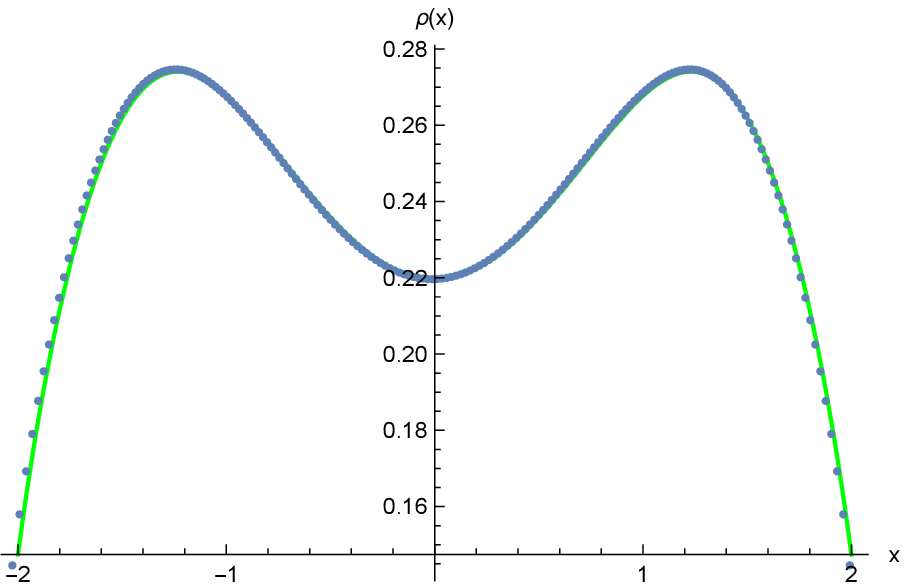}
 \label{nccomp2}
 \end{center}
 \end{minipage}
 \end{tabular}
\caption{These figures show the numerical solution (blue dots) and analytic solutions for $\rho(x)$ (green line). The left utilizes the parameter ($N,N_{f},m$)=(100,2000,2), and the right one is for parameter ($N,N_f,m$)=(100,800,0.5).}
 \end{figure} 

When we take the infinite mass limit, it can naively be thought that this theory becomes a bad theory,  and its matrix model diverges. However, this argument is not correct in the following sense: the density function has peaks around $\pm m$ and the eigenvalues gather around these peaks as $m$ becomes large. Thus, in the large $N$ limit the partition function of this massive SQCD theory corresponds to that of the effective theory on the point of the Coulomb branch where half of the eigenvalues sit on $+m$ and the others sit on $-m$ as 
\begin{align}
\label{Coulomb}
\sigma=\left(
\begin{array}{c|c}
-m\bf{1}_{\frac{N}{2}\times\frac{N}{2}}&\bf{0}\\\hline
\bf{0}&m\bf{1}_{\frac{N}{2}\times\frac{N}{2}}
\end{array}
\right).
\end{align} 

 In fact, this argument is confirmed as follows: We assume that the eigenvalues are separated as 
\begin{align}
x_{i}=\begin{cases}
m-\lambda_{i} \quad (i=1,\dots \frac{N}{2}),\\
-m-\widetilde{\lambda}_{i}\quad  (i=\frac{N}{2}+1\dots N),
\end{cases}
\end{align}
where we assume that $\lambda_{i}$ and $\widetilde{\lambda}_{i}$ do not depend on $m$. The saddle point equations \eqref{sadqcd2} for the first $\frac{N}{2}$ eigenvalues are written as 
\begin{align}
0=&-2\sum_{j\neq i}\coth\pi\left(\lambda_{i}-\lambda_{j}\right)-2\sum_{j}\coth\pi\left(\lambda_{i}-\widetilde{\lambda}_{j}-2m\right)+\frac{N_{f}}{2}\bigg(\tanh\pi\lambda_{i}+\tanh\pi\left(\lambda_{i}-2m\bigg)\right)\nonumber \\
\label{separate1}
\rightarrow 0=&N\left(\frac{N_{f}}{2N}-1\right)+2\sum^{\frac{N}{2}}_{j\neq i}\coth\pi\left(\lambda_{i}-\lambda_{j}\right)-\frac{N_{f}}{2}\tanh\pi\lambda_{i},
\end{align}
where we took the infinite mass limit in the second line and we note that the first term can be interpreted as the gauge-R mixed Chern-Simons term \cite{CDFKS,CDFKS0,CDFK} induced by integrating out the massive gauginos and Majorana fermions of chiral multiplets. For the latter $\frac{N}{2}$ eigenvalues, the saddle point equation in the large mass limit is almost same as \eqref{separate1}:
\begin{align}
\label{separate2}
 0=&N\left(1-\frac{N_{f}}{2N}\right)+2\sum^{\frac{N}{2}}_{j\neq i}\coth\pi\left(\widetilde{\lambda}_{i}-\widetilde{\lambda}_{j}\right)-\frac{N_{f}}{2}\tanh\pi\widetilde{\lambda}_{i}.
\end{align} 
The equations \eqref{separate1} and \eqref{separate2} imply that in the infinite mass limit the matrix model \eqref{matrixqcd2} becomes \footnote{The overall factor of the matrix model cannot be determined in this procedure.}
\begin{align}
\label{effectivematrix}
Z\sim Z_{\text{Massive}}(m)&\int d^\frac{N}{2}\lambda\frac{e^{\pi N\left(\frac{N_{f}}{2N}-1\right)\sum_{i}\lambda_{i}}\prod_{i<j}\left(2\sinh\pi\left(\lambda_{i}-\lambda_{j}\right)\right)^2}{\prod_{i} \left(2\cosh\pi \lambda_{i}\right)^{\frac{N_{f}}{2}}}\nonumber \\
&\times\int d^\frac{N}{2}\widetilde{\lambda}\frac{e^{-\pi N\left(\frac{N_{f}}{2N}-1\right)\sum_{i}\widetilde{\lambda}_{i}}\prod_{i<j}\left(2\sinh\pi\left(\widetilde{\lambda}_{i}-\widetilde{\lambda}_{j}\right)\right)^2}{\prod_{i} \left(2\cosh\pi \widetilde{\lambda}_{i}\right)^{\frac{N_{f}}{2}}},
\end{align} 
because the saddle point equation of this is equivalent to \eqref{separate1} and \eqref{separate2}. The factor $Z_{\text{Massive}}(m)$ represents the contribution of the decoupled free massive degrees of freedom. We can evaluate $Z_{\text{Massive}}\sim M^{-\frac{N}{2}(N_f-N)}$. This part cannot be obtained from the saddle point equations. This represents SQCD theories with the two U($\frac{N}{2}$) gauge group, $\frac{N_{f}}{2}$ fundamental hypermultiplets, and an FI parameter $\pm N\left(1-\frac{N_{f}}{2N}\right)$. \footnote{
To be precise, the FI parameter is given by $\frac{1}{r_{{S}^3}}\left(1-\frac{N_{f}}{2N}\right)$ if we recover the radius of $S^{3}$ because in a 3d theory an FI parameter has mass dimension 1.}
 As previously noted, the FI term is induced by one-loop effects as a mixed Chern-Simons term consisting of vector multiplets of the gauge and R-symmetry by integrating out the effectively massive fermions. We argue on this point in Appendix \ref{mix}. This FI term cannot appear when we consider gauge on theories on flat space.

In fact, we verify our assumptions by comparing the density function of the matrix model of the effective theory \eqref{effectivematrix} with that of the matrix model \eqref{matrixqcd2} in the infinite mass limit. First, we consider the density function of SQCD with massive hypermultiplets \eqref{densitysqcd3} in the infinite mass limit. We rewrite $X$ as $X=MZ$ and assume that $Z$ is order $\mathcal{O}(M^0)$. This procedure corresponds to simultaneous shifting $x_{i}$ by $m$ and focusing on the peak of the density function around $+m$. We have to consider the expansion of $a$ \eqref{interbal1} around $m=\infty$, which is given by
\begin{align}
a= \alpha M+\mathcal{O}(M^{0}),\quad \alpha\equiv\frac{4(\xi-1)}{(\xi-2)^2}.
\end{align} 
Thus, the density function is expanded around $m=\infty$ as
\begin{align}
\label{densityinfty}
\rho(x)= \frac{\xi}{2 (Z+1)}\sqrt{\frac{Z\left(\alpha-Z\right)}{1+\alpha}}+\mathcal{O}(M^{-1}),
\end{align}
where $Z \in [0,\alpha]$ in the infinite mass limit. Then, we compare this with the solution for the saddle point equation of the $\lambda$ part \eqref{separate1} because the $\lambda$ part corresponds to a part of the massive SQCD in which the eigenvalues are concentrated on $+m$. The solution of its saddle point equation \eqref{separate1} is given by applying the result in Section \ref{FI}. In this case, $a$ and $b$ are 
\begin{align}
a=\alpha,\quad b=0,
\end{align}
and the density function is
\begin{align}
\label{densityinfty2}
\rho(z)=\frac{\xi}{2(Z+1)}\sqrt{\frac{Z\left(\alpha-Z\right)}{1+\alpha}}, \quad Z\equiv e^{2\pi z},
\end{align}
where the additional factor of $\frac{1}{2}$ results from the fact that the effective theory has two U($\frac{N}{2}$) gauge groups and the normalization condition should be taken as
\begin{align}
\int_{I}\frac{dZ}{2\pi Z}\rho(z)=\frac{1}{2}.
\end{align}
The density functions \eqref{densityinfty} and \eqref{densityinfty2} are completely equivalent. Next, we should consider the part concentrated around $-m$. Here, we must rewrite $X=M^{-1}Z$ in \eqref{densitysqcd3} and the density function in this limit is
\begin{align}
\label{densityinfinity3}
\rho(x)=\frac{\xi}{2(Z+1)}\sqrt{\frac{Z-\frac{1}{\alpha}}{1+\frac{1}{\alpha}}}+\mathcal{O}(M^{-1}),
\end{align}
where $Z\in [\frac{1}{\alpha},\infty]$. Then, we consider $\widetilde{\lambda}$ part of \eqref{effectivematrix}. The solution of its saddle point equation is given by applying the result of \ref{FI} to \eqref{separate2} and we obtain 
\begin{align}
a=\infty,\quad b=\frac{1}{\alpha},
\end{align}
with the density function 
\begin{align}
\rho(z)=\frac{\xi}{2(Z+1)}\sqrt{\frac{Z-\frac{1}{\alpha} }{1+\frac{1}{\alpha}}}.
\end{align}
This is the same as \eqref{densityinfinity3}. Therefore, we conclude that SQCD with $N_{f}$ massive hypermultiplets, as studied here, becomes two SQCDs in the infinite mass limit: each is a U($\frac{N}{2}$) SQCD with $\frac{N_f}{2}$ massless hypermultiplets and the FI term $\zeta=\pm iN(\frac{2N_{f}}{N}-1)$. This result suggests that if we assume that the massive matter fields will be decoupled, then the sphere partition function of a  massive theory that would become a bad theory always becomes that of a specific effective theory. This means that an interacting SCFT on a specific singular point of the Coulomb branch appears in the infinite mass limit, rather than  a bad theory appearing. This result may also suggest that the massive theory cannot be employed for the UV regularization of the bad theory. In Section \ref{finite}, we will verify our claim through the exact calculation of the partition function of finite-rank SQCD. It is expected that the partition function can be written as the product of those of the sector of the decoupled free massive multiplets and of the effective theory in the infinite mass limit.

\subsection{SQCD with massive and massless hypermultiplets}
\label{33}
In the previous subsection, all matter fields of the theory were set to be massive. In this subsection, we consider an SQCD theory with both massive and massless matter fields. It is expected that the asymptotic behavior of the partition function in the infinite mass limit will depend on the number of massless matter fields because the presence of the sufficient number of matter fields makes the matrix model convergent.

We consider U($N$) SQCD with $\frac{N_{f}}{3}$ pairs of  massive hypermultiplets with $\pm m$  and $\frac{N_{f}}{3}$ massless hyper multiplets. We assume that $\frac{N_{f}}{3}$ is an integer. The matrix model is given by 
\begin{align}
\label{matrixqcd3}
Z=\frac{1}{N!}\int \prod_{i=1}^{N}dx_{i}\frac{\prod_{i<j}4\sinh^2\left(\pi (x_{i}-x_{j})\right)}{\prod_{i}\bigg(2\cosh\pi (x_{i}+m)2\cosh\pi (x_{i}-m)2\cosh\pi(x_{i})\bigg)^{\frac{N_{f}}{3}}}.
\end{align}
The saddle point equation is given by 
\begin{align}
2\sum_{j\neq i}^{N}\coth\pi(x_{i}-x_{j})=\frac{N_{f}}{3}\left(\tanh\pi\left(x_{i}+m\right)+\tanh\pi\left(x_{i}-m\right)+\tanh\pi x_{i}\right),
\end{align}
and we take the continuous limit of this. This is written as 
\begin{align}
6\left(\text{P}\int_{\mathcal{C}} dy\coth\pi(x-y)\right)=\xi\bigg(\tanh\pi(x+m)+\tanh\pi(x-m)+\tanh\pi x\bigg).
\end{align}
We define the resolvent $\omega(X)$ and potential $V^{\prime}(x)$ as
\begin{align}
\omega(X)=&6 \int dy \rho(y)\frac{X+Y}{X-Y}=6\bigg(1+\int \frac{dY}{\pi}\frac{\rho(y)}{X-Y}\bigg),\\
V^{\prime}(x)=&\xi\left(\frac{X-1}{X+1}+\frac{X-M}{X+M}+\frac{X-M^{-1}}{X+M^{-1}}\right),
\end{align} 
where $M=e^{2\pi m}$. The resolvent is obtained from \eqref{omega1} as
\begin{align}
\omega(X)=\omega_{0}(X;1;a,\frac{1}{a})+\omega_{0}(X;M;a,\frac{1}{a})+\omega_{0}(X;M^{-1};a,\frac{1}{a}).
\end{align}
The cut $\mathcal{C}=[\frac{1}{a},a]$ is determined by the following asymptotic equation:
\begin{align}
\label{asymde}
-\frac{6}{\xi}=-3+\frac{2}{\sqrt{(1+a)(1+\frac{1}{a})}}+\frac{2}{\sqrt{(M+a)(M+\frac{1}{a})}}+\frac{2}{\sqrt{(M^{-1}+a)(M^{-1}+\frac{1}{a})}}.
\end{align}
Unfortunately, there are generally no explicit forms of the solution because this equation corresponds to an octic equation in $a$ . However, we can determine the solution numerically or in the infinite mass limit.  The density function for this case is given by \eqref{analy2} as
\begin{align}
\label{density2}
\rho(x)
=&\frac{\xi}{3}\bigg[\frac{M\sqrt{(a-X)(X-\frac{1}{a})}}{(X+M)\sqrt{(M+a)(M+\frac{1}{a})}}+\frac{M^{-1}\sqrt{(a-X)(X-\frac{1}{a})}}{(X+M^{-1})\sqrt{(M^{-1}+a)(M^{-1}+\frac{1}{a})}}\nonumber \\
&\quad\quad\quad+\frac{\sqrt{(a-X)(X-\frac{1}{a})}}{(X+1)\sqrt{(1+a)(1+\frac{1}{a})}}\bigg].
\end{align}
Let us consider what happens to the matrix model when the number of the matter fields varies. When $\frac{N_f}{3}\geq2N$, the matrix model is still well defined after we take the infinite mass limit and all massive matter fields decouple from the theory. In fact, in this case the limit in which the mass is  taken to infinity and the integrals of the matrix model are commutative \footnote{
In this work, we focus only on the leading part of the mass infinite limit. Namely, when there exist finite constants $\alpha$ and $\beta$ such that the relation 
\begin{align}
\lim_{M\to \infty} \left(\int_{-\infty}^{\infty}dx f(x,M)M^{\alpha}\right)=\int_{-\infty}^{\infty}dx \lim_{M\to \infty}\left(f(x,M)M^{\alpha}\right)=\beta,
\end{align}
is satisfied for $f(x,M)$, which is a function of $x$ and $M$, we say that the infinite integral and the limit of $M$ are commutative. 
}. This immediately implies that all the massive matter fields decouple and the remaining theory is U($N$) SQCD with $\frac{N_{f}}{3}$ massless matter fields. This situation is reflected in the equation \eqref{asymde}. We assume that the solution does not depend on $M$\footnote{This assumption means that the effectively massless degrees of freedom cannot appear.
} when we take the mass $m$ to infinity. Then, the equation \eqref{asymde} becomes the same as \eqref{xi} for the case with $\frac{N_f}{3}$ flavors.
\begin{align}
3-\frac{6}{\xi}=\frac{2}{\sqrt{(1+a)(1+\frac{1}{a})}}.
\end{align}
This implies that the solution of \eqref{asymde}, which does not depend on the mass $m$ can exist when $\frac{N_{f}}{3}\geq 2N$ while the constant solution cannot exist in the infinite mass limit when $\frac{N_{f}}{3}<2N$. The numerical analysis of \eqref{asymde} supports the existence of such a solution. Indeed, the density function is the same as that of U($N$) gauge theory with $\frac{N_{f}}{3}$ massless hypermultiplets and this means that all the massive hypermultiplets decouple from the theory because in the infinite mass limit, the origin of the Coulomb branch is dominant.

\begin{figure}[h!]
 \begin{minipage}{0.5\hsize}
 \begin{center}
 \includegraphics[width=8cm]{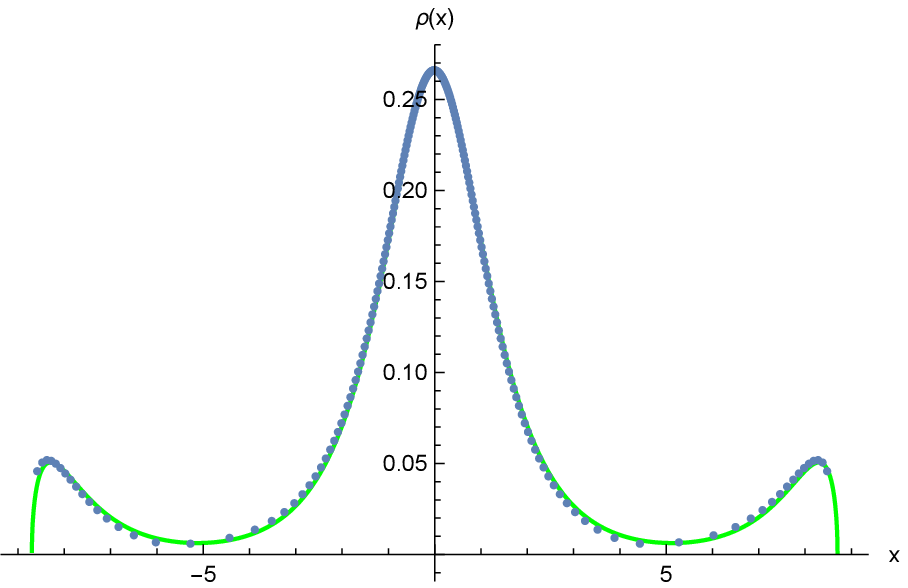}
 \end{center}
 \end{minipage}
  \begin{minipage}{0.5\hsize}
 \begin{center}
 \includegraphics[width=8cm]{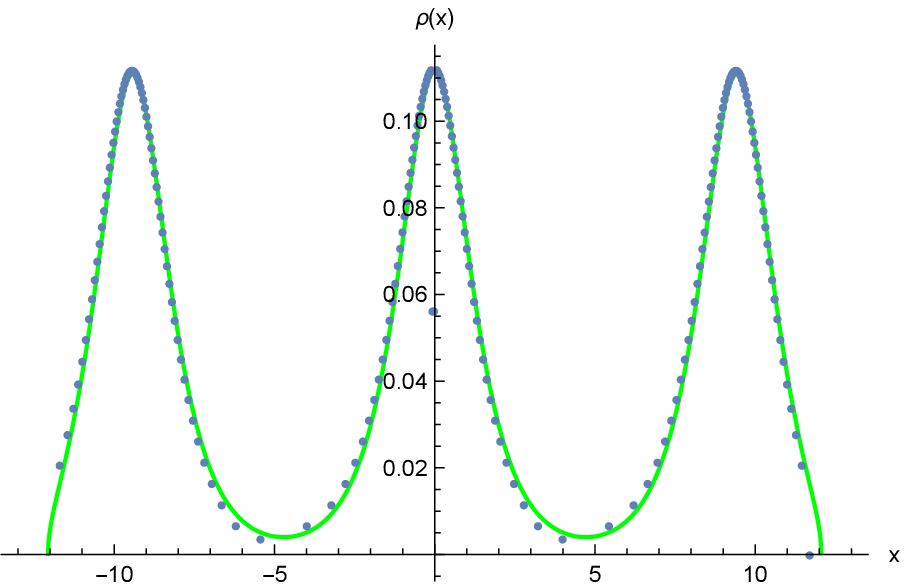}
 \end{center}
 \end{minipage}
 \caption{These figure shows the density function $\rho(x)$ \eqref{density2} (green line) and the numerical one from the saddle point equation (blue dots). The left and right figures correspond to ($N$,$N_{f}$,$m$)=(200,1000,3) and (200,420,3) respectively.}
  \label{density3}
 \end{figure}

 On the other hand, when $\frac{N_{f}}{3}< 2N$, $a$ is proportional to $M$ in the infinite mass limit, and the density function has three peaks: around the origin and $x=\pm m$. We illustrate the behavior of the density function $\rho(x)$ in Figure \ref{density3}.
 Then, we study the effective theory that appears in this situation by analyzing the behavior of the density function when we take the infinite mass limit. First, we need to know how the gauge group U($N$) is broken. From the density function, we find that U($N$) is broken into three parts. Thus, we assume that
\begin{align}
\text{U}(N)\rightarrow \text{U}(N_{1})\times\text{U}(N_{2})\times \text{U}(N_{3}),\quad (N_{1}+N_{2}+N_{3}=N).
\end{align}
The rank of each of the three gauge groups is determined by the ratio of the numbers of eigenvalues around each peak. The density function $\rho(x)$ counts the number of the eigenvalues between $x$ and $x+dx$. Therefore, we count the numbers of the eigenvalues that exist around each peak by integrating the corresponding density function in the infinite mass limit to determine $N_{1}$, $N_{2}$ and $N_{3}$.

We assume that there exists a solution proportional to $M$. In the infinite mass limit, the equation \eqref{asymde} becomes
 \begin{align}
 1-\frac{2}{\xi}=\frac{2}{3\sqrt{(1+\beta)}},
 \end{align}
 where we assume that $a=M\beta$. We can then immediately determine $\beta$ as 
 \begin{align}
 \beta=\frac{(5\xi-6)(-\xi+6)}{9(\xi-2)^2}.
 \end{align}
 In order to study the behavior of the density function around $x=m$, we redefine $X$ using an order $\mathcal{O}(M^0)$ variable $Z$ as $X=MZ$ and take $M\rightarrow \infty$. In this limit, the density function \eqref{density2} becomes
\begin{align}
\label{peak1}
\rho(x)\xrightarrow[m \to \infty]{} \frac{\xi}{3(Z+1)}\sqrt{\frac{Z(\beta-Z)}{1+\beta}}\equiv \rho_{+}(z), \quad Z \equiv e^{2\pi z},
\end{align}
 where $Z \in [0,\beta]$. Next, we examine the density function around the peak at $x=-m$ by regarding $X$ as $X=M^{-1}Z$ in \eqref{density2}. By the same calculation as in \eqref{densityinfty2}, this becoms
 \begin{align}
 \label{peak2}
 \rho(x) \xrightarrow[m \to \infty]{} \frac{\xi}{3(Z+1)}\sqrt{\frac{Z-\frac{1}{\beta}}{1+\frac{1}{\beta}}}\equiv \rho_{-}(z),
 \end{align}
where $Z\in [\frac{1}{\beta},\infty]$. The final part is the density function around $x=0$. To investigate this part of the density function, we assume that $X$ is of order $\mathcal{O}(M^{0})$. Then, we take the infinite mass limit and the density function becomes 
\begin{align}
\label{peak3}
\rho(x)\xrightarrow[m \to \infty]{} \frac{\xi\sqrt{Z}}{3(Z+1)}\equiv \rho_{0}(z),
\end{align}
where $Z$ takes value $\in$ $[0,\infty]$.
To determine $N_{1}$, $N_{2}$ and $N_{3}$, we integrate \eqref{peak1}, \eqref{peak2} and \eqref{peak3}, respectively. We obtain
\begin{align}
\int_{0}^{\beta}\frac{dZ}{2\pi Z}\rho_{+}(z)&=\frac{6-\xi}{12},\\
\int_{\frac{1}{\beta}}^{\infty}\frac{dZ}{2\pi Z}\rho_{-}(z)&=\frac{6-\xi}{12},\\
\int_{0}^{\infty}\frac{dZ}{2\pi Z}\rho_{0}(z)&=\frac{\xi}{6}.
\end{align}
This result implies that the gauge group U($N$) is broken into the following:
\begin{align}
\label{rank}
N_{1}=\frac{\xi}{6}N,\quad N_{2}=N_{3}=\frac{6-\xi}{12}N,
\end{align}
where we assume that $\frac{\xi}{6}N$ and $\frac{6-\xi}{12}N$ are integers. This implies that in the infinite mass limit, the theory becomes the effective theory on a point of the Coulomb branch as 
\begin{align}
\label{Coulomb2}
\sigma=\left(
\begin{array}{c|c|c}
-m\bf{1}_{N_{2}\times N_{2}}& &\\ \hline
&\bf{0}_{N_1\times N_{1}}& \\ \hline
& &  m\bf{1}_{N_2\times N_2}
\end{array}
\right).
\end{align}

We assume that the eigenvalues are separated as 
\begin{align}
x_{i}=\begin{cases}
-m-\lambda^{1}_{i},\quad (i=1,\dots,N_{2}),\\
\lambda^{2}_{i},\quad (i=N_{2}+1,\dots N_{1}+N_{2}),\\
m-\lambda^{3}_{i} \quad (i=N_{1}+N_{2}+1,\dots, N).
\end{cases}
\end{align}
Through a similar calculation as that presented in the previous subsection, the saddle point equation is rewritten in the following three parts:
\begin{align}
\label{effsaddle1}
0=&2N\left(\frac{6+\xi}{12}-\frac{\xi}{3}\right)+2\sum_{j\neq i}^{N_{2}}\coth\pi\left(\lambda^1_{i}-\lambda^{1}_{j}\right)-\frac{N_{f}}{3}\tanh\pi\lambda^1_{i},\quad (i=1,\dots N_{2}),\\
\label{effsaddle2}
0=&2\sum_{j\neq i}^{N_{1}}\coth\pi\left(\lambda^2_{i}-\lambda^{2}_{j}\right)-\frac{N_{f}}{3}\tanh\pi\lambda^2_{i},\quad (i=1,\dots N_{1})\\
\label{effsaddle3}
0=&-2N\left(\frac{6+\xi}{12}-\frac{\xi}{3}\right)+2\sum_{j\neq i}^{N_{2}}\coth\pi\left(\lambda^3_{i}-\lambda^{3}_{j}\right)-\frac{N_{f}}{3}\tanh\pi\lambda^3_{i},\quad (i=1,\dots N_{2}).
\end{align}
These equations mean that the matrix model \eqref{matrixqcd2} in the infinite mass limit becomes the following matrix model:
\begin{align}
\label{effectivematrix2}
Z=&Z_{\text{massive}}(m)\int d^{N_{2}}\lambda^2\frac{e^{2\pi N\left(\frac{\xi}{3}-\frac{6+\xi}{12}\right)\sum_{i}\lambda^2_{i}}\prod_{i<j}\left(2\sinh\pi\left(\lambda^2_{i}-\lambda^2_{j}\right)\right)^2}{\prod_{i} \left(2\cosh\pi \lambda^2_{i}\right)^{\frac{N_{f}}{3}}}\nonumber \\
&\times \int d^{N_{2}}\lambda^{3}\frac{e^{-2\pi N\left(\frac{\xi}{3}-\frac{6+\xi}{12}\right)\sum_{i}\lambda^3_{i}}\prod_{i<j}\left(2\sinh\pi\left(\lambda^3_{i}-\lambda^3_{j}\right)\right)^2}{\prod_{i} \left(2\cosh\pi \lambda^3_{i}\right)^{\frac{N_{f}}{3}}}\int d^{N_{1}}\lambda^1\frac{\prod_{i<j}\left(2\sinh\pi\left(\lambda^1_{i}-\lambda^1_{j}\right)\right)^2}{\prod_{i}\left(2\cosh\pi \lambda^1_{i}\right)^{\frac{N_{f}}{3}}}.
\end{align} 
The two U($N_{2}$) parts have FI terms, which also arise from the gauge-R-symmetry mixed Chern-Simons term we discussed in the previous section. The U($N_{1}$) part has no FI terms since there are pairs of mixed Chern-Simons terms which have opposite overall signs corresponding to those of the masses of the effectively massive fermions. The decoupled massive free sector can be estimated by $Z_{\text{Massive}}(m) \sim M^{-\frac{N_f N(6+\xi)}{36}}$.

In fact, we can confirm that the density functions obtained in the infinite mass limit are the same as those obtained from \eqref{effsaddle1}, \eqref{effsaddle2} and \eqref{effsaddle3}. First, the solution of \eqref{effsaddle1} is given by \eqref{rho1} and \eqref{length1}. We obtain 
\begin{align}
a=\infty, \quad \rho(z)=\frac{\xi\sqrt{Z}}{3\pi(Z+1)},
\end{align}
where $Z \in [0,\infty]$ and we assume that $\frac{Z}{a}$ is zero since when we scale $Z=a\tilde{Z}$, $\rho(z)$ is $\mathcal{O}(\frac{1}{a})$ and only the order $\mathcal{O}(a^{0})$ part of $Z$ can contribute to $\rho(z)$. This density function is same as $\rho_{0}(z)$. Next, we consider the solution of \eqref{effsaddle2}. We can obtain the solution from the equations \eqref{soleta1} as 
\begin{align}
a=\infty ,\quad b=\frac{(\tilde{\xi}-2)^2}{4(\tilde{\xi}-1)}=\frac{1}{\beta},\quad \tilde{\xi}\equiv \frac{4\xi}{6-\xi},
\end{align}
and the density function is given by 
\begin{align}
\rho(z)=\frac{6-\xi}{12}\frac{\tilde{\xi}}{(Z+1)}\sqrt{\frac{Z-\frac{1}{\beta}}{1+\frac{1}{\beta}}}=\frac{\xi}{3(Z+1)}\sqrt{\frac{Z-\frac{1}{\beta}}{(1+\frac{1}{\beta})}},
\end{align}
where $Z\in [\frac{1}{\beta},\infty]$. This corresponds to $\rho_{-}(z)$. Finally, we solve \eqref{effsaddle3}. Its solution is obtained in the same manner as that of \eqref{effsaddle2}. The solution is given as
\begin{align}
a=\frac{4(\tilde{\xi}-1)}{(\tilde{\xi}-2)^2}=\beta,\quad b=0,
\end{align}
and 
\begin{align}
\rho(z)=\frac{6-\xi}{12}\frac{\tilde{\xi}}{(Z+1)}\sqrt{\frac{Z(\beta-Z)}{1+\beta}}=\frac{\xi}{3(Z+1)}\sqrt{\frac{Z(\beta-Z)}{1+\beta}}.
\end{align}
This density function is same as $\rho_+(z)$. In the above calculation, the normalization condition of each density function is set such that they corresponds to each rank of the gauge groups \eqref{rank}. We conclude that the matrix model \eqref{matrixqcd3} becomes the matrix model \eqref{effectivematrix2} when we take the infinite mass limit. The above result shows that the massive multiplets cannot decouple from the theory, and so the matrix model of the theory converges. In other words, a ``good" theory cannot become a ``bad" theory after the massive matter fields decouple although the matrix model can become a ``good" theory after the decoupling of the massive matter multiplets. A notable result is that the gauge group of the effective theory depends on the number of flavors. When $\xi=2$, the effective theory consist of three SQCDs with gauge groups U($\frac{N}{3}$) with $\frac{N_{f}}{3}$ hypermultiplets and the gauge group is maximally broken between $2\leq \xi \leq 6$. Then, the gauge group recovers to U($N$) as $\xi$ increases. We note that the matrix model of the effective theory when $2\leq \xi \leq 6$ is always convergent.

\section{Finite rank SQCD}
\label{finite}

In SQCD cases, the matrix model can be actually calculated at least for a sufficiently low rank of the gauge group. In this section, we confirm through the exact results that our argument for the effective theory is true even in case of finite $N$. Furthermore, we show what happens in the infinite mass limit for theories that are not covered by our argument in the large $N$ limit.

\subsection{With massive hypermultiplets}

\subsubsection*{U(1) SQED}
The matrix model of SQED with massive matter fields is given by 
\begin{align}
Z_{\text{U}(1)}^{N_{f}}=\int_{-\infty}^{\infty}dx\frac{1}{\left(2\cosh\pi(x+m)2\cosh\pi(x-m)\right)^{\frac{N_{f}}{2}}}.
\end{align}
This model is considered in \cite{RT} with an FI term. In this case, the theory may not become the effective theory expected from the previous section because $\frac{N}{2}$ is not integer. Thus, we want to know what happens in this case when $m\rightarrow \infty$. The exact result for any $N_{f}$ is written in terms of the hypergeometric function as \cite{RT}
\begin{align}
Z^{N_f}_{\text{U}(1)}=\frac{\Gamma(\frac{N_f}{2})}{2^{\frac{N_f}{2}}\sqrt{2\pi}\Gamma(\frac{N_f}{2}+\frac{1}{2})\left(\cosh2\pi m+1\right)^{\frac{N_f}{2}-\frac{1}{2}}}\  _{2}F_{1}\left(\frac{1}{2},\frac{1}{2},\frac{N_f}{2}+\frac{1}{2};\frac{1}{2}(1-\cosh2\pi m)\right).
\end{align}
The leading part in the infinite mass limit of this partition function is given by
\begin{align}
Z^{N_{f}}_{\text{U}(1)} \xrightarrow[m \rightarrow \infty]{} \frac{ 1}{\pi}\frac{\log M}{M^{\frac{N_f}{2}}},\quad M\equiv e^{2\pi m}.
\end{align}
This is a strange result in the sense of our argument so far because there cannot exist a $\log M$ term when a massive theory splits into a decoupled sector and an SCFT sector in the mass infinite limit. Therefore, we cannot determine what the effective theory is in this case using our previous argument.

\subsubsection*{U(2) SQCD}
\label{u2sqcd}
The partition function for this theory is given by 
\begin{align}
Z^{N_{f}}_{\text{U}(2)}=\frac{1}{2!}\int^{\infty}_{-\infty}dx\int_{-\infty}^{\infty} dy\frac{4\sinh^2\pi(x-y)}{\left(2\cosh\pi(x+m)2\cosh\pi(x-m)2\cosh\pi(y+m)2\cosh\pi(y-m)\right)^{\frac{N_{f}}{2}}},
\end{align}
and the results for small $N_{f}$ are summarized in the following table.

\begin{table}[ht!]
  \begin{center}
  \large
    \begin{tabular}{|c|c|c|c|} \hline
          $ N_f=4$   &  $N_{f}=6$&   $N_{f}=8$&$ N_{f}=10$\\ \hline
          $\frac{1}{(2\pi)^2M^2}$&$\frac{1}{(4\pi)^2 M^4}$ &$\frac{1}{(6\pi)^2M^6} $& $\frac{1}{(8\pi)^2M^8}$     \\ \hline 
    \end{tabular}
      \caption{The leading part of $Z^{N_f}_{\text{U}(2)}$ when $m\rightarrow \infty$.}
  \end{center}
\end{table}

In fact, these results show that in the infinite mass limit, the partition function can be interpreted as that of the theory expected from the large $N$ calculation since the following relation is verified:  
\begin{align}
Z_{\text{Massive}}Z_{\text{U}(1)\times \text{U}(1)}&=\frac{4\sinh^2(2\pi m)}{\left(2\cosh 2\pi m\right)^{N_f}}\int^{\infty}_{-\infty}dx\frac{e^{\pi\left(2-\frac{N_f}{2}\right)x}}{(2\cosh\pi x)^{\frac{N_f}{2}}}\int^{\infty}_{-\infty}dx\frac{e^{-\pi\left(2-\frac{N_f}{2}\right)x}}{(2\cosh\pi x)^{\frac{N_{f}}{2}}}\nonumber \\
&\xrightarrow[m \to \infty]{} \frac{1}{\left((N_{f}-2)\pi\right)^2M^{(N_{f}-2)}},
\end{align}
where the two integrals represent a U(1)$\times$ U(1) theory and the pre-factor $Z_{\text{massive}}$ represents decoupled massive free sector, for which the denominator comes from the massive hypermultiplets and the numerator comes from the vector multiplets. Because $\frac{N}{2}$ is an integer, $\U(2)$ can be broken down $\U(1) \times \U(1)$. In this case, we see that the partition functions are equal including the overall factor, which can not be determined from the large $N$ analysis.

\subsubsection*{U(3) SQCD}
The partition function for this case is given by 
\begin{align}
Z^{N_f}_{\text{U}(3)}=\frac{1}{3!}\int_{-\infty}^{\infty}dxdydz\frac{4\sinh^2\pi(x-y)4\sinh^2\pi(x-z)4\sinh^2\pi(y-z)}{\left(2\cosh\pi(x\pm m)2\cosh\pi(y\pm m)2\cosh\pi(z\pm m)\right)^{\frac{N_f}{2}}},
\end{align}
where we define 
\begin{align}
2\cosh\pi(X\pm Y)\equiv 2\cosh\pi(X+Y)2\cosh\pi(X-Y).
\end{align}
The results for small $N_{f}$ are given by the following table.

\begin{table}[ht!]
  \begin{center}
  \large
    \begin{tabular}{|c|c|c|c|} \hline
          $ N_f=6$   &  $N_{f}=8$&   $N_{f}=10$&$ N_{f}=12$\\ \hline
          $\frac{\log M }{16\pi^3 M^5}$&$\frac{\log M}{144 \pi^3 M^8}$ &$\frac{\log M}{576\pi^3 M^{11}} $& $\frac{\log M}{1600 \pi^3M^{14}}$     \\ \hline 
    \end{tabular}
      \caption{The leading part of $Z^{N_f}_{\text{U}(3)}$ when $m \rightarrow \infty$}
  \end{center}
\end{table}
From these results, it may not be possible that the effective theory is composed of an SCFT and a massive free sector for the same reason as in the U(1) case, namely, that $\frac{N}{2}$ is not an integer.

\subsection{With massive and massless hypermultiplets}

\subsubsection*{U(1) SQED}
This case is trivial because the limit that takes mass to infinity is commutative with the infinite integral. The massive matter fields simply decouple from the theory and the remaining theory is SQED with $\frac{N_{f}}{3}$ massless hypermultiplets. In fact, 
\begin{align}
\widetilde{Z}^{N_f}_{\U(1)}&=\int_{-\infty}^{\infty}dx\frac{1}{\left(2\cosh\pi x2\cosh\pi(x+m)2\cosh\pi(x-m)\right)^{\frac{N_{f}}{3}}}\nonumber \xrightarrow[m \to \infty]{} \left(\frac{1}{M}\right)^{\frac{N_f}{3}}\widetilde{Z}^{\frac{N_f}{3}}_{\U(1)}\big|_{m=0},
\end{align}
where for $\mathcal{N}=4$ U($N$) SQCD with a massless flavors part, $\widetilde{Z}^{N_f}_{\U(N)}\big|_{m=0}$ can be calculated for $N_f \geq 2N$ \cite{Tierz,YB} as
\begin{align}
\label{fQCD}
\widetilde{Z}^{N_f}_{\U(N)}\big|_{m=0}=\frac{1}{N!}\frac{1}{(2\pi)^N}\prod_{k=0}^{N-1} \frac{\Gamma(k+2)\left(\Gamma(\frac{N_f}{2}-N+k+1)\right)^2}{\Gamma(N_f-N+k+1)}.
\end{align}

\subsubsection*{U(2) SQCD}

In the following, we present the exact calculation of the partition function of U(2) SQCD with $N_{f}$ fundamental hypermultiplets,
\begin{align}
\widetilde{Z}^{N_{f}}_{\U(2)}=\frac{1}{2!}\int^{\infty}_{-\infty} dxdy \frac{4\sinh^2\pi(x-y)}{\left(2\cosh\pi(x\pm m)2\cosh(y\pm m)2\cosh\pi x2\cosh\pi y\right)^{\frac{N_{f}}{3}}}.
\end{align}
The results for small $N_{f}$ are summarized in following table.

\begin{table}[ht!]
  \begin{center}
  \large
    \begin{tabular}{|c|c|c|c|} \hline
          $ N_f=3$   &  $N_{f}=6$&   $N_{f}=9$&$ N_{f}=12$\\ \hline
        $\frac{1}{4 M}$&$\frac{\left(\log M\right)^2}{4\pi^2 M^4}$ &$\frac{1}{32 M^{6}} $& $\frac{1}{48 \pi^2 M^{8}}$     \\ \hline 
    \end{tabular}
      \caption{The leading part of $\tilde{Z}^{N_{f}}_{U(2)}$ when $m\rightarrow \infty$.}
  \end{center}
\end{table}

When $N_{f}=3$, we can deduce the effective theory as follows:
\begin{align}
Z_{\text{Massive}}Z_{\U(1)\times \U(1)}= \frac{4\sinh^2 2\pi m}{\left(2\cosh 2\pi m 2\cosh\pi m\right)^2}\left(\int_{-\infty}^{\infty}dx\frac{1}{2\cosh\pi x}\right)^2\xrightarrow[m \to \infty]{} \frac{1}{4M}.
\end{align}
This means that the effective theory appears when we chose the point of the Coulomb branch as
\begin{align}
\sigma=\left(
\begin{array}{cc}
-m& \\
&m
\end{array}
\right),
\end{align}
in the sense of theories on the flat space. Because our previous expectation cannot be applied to this case as $\frac{\xi}{6}N$ and $\frac{12-\xi}{12}N$ are not integers, it may be expected that a $\log M$ term also appears in the infinite mass limit as in the case with only massive matter fields. However, a $\log M$ term does not appear in this case and the effective theory is the expected one.
When $N_f=6$, a $\log M$ term does appear.  In this case, the effective theory may not be U(1)$\times$U(1). It is also notable that whether or not a $\log M$ appears depends on the number of flavors.
When $N_f\geq9$, the infinite mass limit is commutative with the integral. Thus, the result is trivial \footnote{Exactly speaking, the matrix converges when $2N-2\leq\frac{N_{f}}{3}$. In the large $N$ limit the order one part is neglected.}.
This is consistent with the fact that the $\frac{2NN_{f}}{3}$ massive matter fields with the real mass $m$ simply decouple by choosing the origin of the Coulomb branch since the remaining theory is a good theory. Namely, in the infinite mass limit, the integral is written as
\begin{align}
\widetilde{Z}^{N_{f}}_{\U(2)}\xrightarrow[m \to \infty]{} \left(\frac{1}{M}\right)^{\frac{2N_{f}}{3}}\widetilde{Z}^{\frac{N_f}{3}}_{\U(2)}\big|_{m=0}.
\end{align}

\subsubsection*{U(3) SQCD}
For this case, the partition function is give by 
\begin{align}
\tilde{Z}^{N_{f}}_{\text{U}(3)}=\frac{1}{3!}\int_{-\infty}^{\infty}\frac{4\sinh^2\pi(x-y)4\sinh^2\pi(x-z)4\sinh^2\pi(y-z)dxdydz}{\left(2\cosh(\pi x) 2\cosh(\pi y)  2\cosh(\pi z)   2\cosh\pi(x\pm m)2\cosh\pi(y\pm m)2\cosh\pi(z\pm m)\right)^{\frac{N_{f}}{3}}}.
\end{align}
The results for small $N_f$ are summarized in the following table.

\begin{table}[ht!]
  \begin{center}
  \large
    \begin{tabular}{|c|c|c|c|} \hline
          $ N_f=6$   &  $N_{f}=9$&   $N_{f}=12$&$ N_{f}=15$\\ \hline
          $\frac{1}{(2\pi)^3 M^4}$&$\frac{9}{2^{12}M^8}$ &$\frac{\left(\log M\right)^2}{48\pi^3 M^{12}} $& $\frac{1}{2^{13}M^{15}}$     \\ \hline 
    \end{tabular}
      \caption{The leading part of $\tilde{Z}^{N_f}_{\U(3)}$ when $m\rightarrow\infty$.}
  \end{center}
\end{table}

In the $N_{f}=6$ case, the theory will become a U(1)$\times$U(1)$\times$U(1) gauge theory, where each gauge group has two massless fundamental hypermultiplets. This theory is expected from the large $N$ analysis when  $\xi=2$ and $\frac{\xi}{6}N$ and $\frac{12-\xi}{6}N$ are integers. In fact, its matrix model is given by
\begin{align}
Z_{\text{Massive}}Z_{\U(1)\times\U(1)\times \U(1)}&=\frac{(4\sinh^2 \pi m)^24\sinh^22\pi m}{(2\cosh\pi m)^4(2\cosh 2\pi m)^6}\left(\int_{-\infty}^{\infty}dx\frac{1}{\left(2\cosh\pi x \right)^2}\right)^3\nonumber \\
&\xrightarrow[m \to \infty]{} \frac{1}{(2\pi)^3M^4}.
\end{align}
When $N_{f}=9$, this means that $\xi=3$. However, $\frac{\xi}{6}N$ and $\frac{12-\xi}{12}N$ are not integers. Therefore, we cannot apply the result from the large $N$ analysis to this case. However, we can guess that the effective theory will be a U(1)$\times$ U(1) $\times$ U(1) gauge theory where each group has three massless hypermultiplets. As opposed to the $N_{f}=6$ case, two of the three U(1) have an imaginary FI term which arises from one-loop effects. In fact, the matrix model is 
\begin{align}
Z_{\text{massive}}Z_{\U(1)\times\U(1)\times \U(1)}&=\frac{\left(4\sinh^2\pi m \right)^2 \left(4\sinh^2 2\pi m\right)}{(2\cosh2\pi m)^6(2\cosh\pi m)^{12}}\left(\int_{-\infty}^{\infty}dx\frac{e^{-2\pi x}}{(2\cosh\pi x)^3}\right)^2\int_{-\infty}^{\infty}dx\frac{1}{(2\cosh\pi x)^3 }\nonumber \\
&\xrightarrow[m \to \infty]{} \frac{9}{2^{12}M^8},
\end{align}
which is the same as $\tilde{Z}^{N_f=9}_{\U(3)}$ in the infinite mass limit. Thus, in the case of $N_{f}=6,9$, we conclude that the IR effective theory corresponds to the theory on a non-trivial Coulomb branch point 
\begin{align}
\sigma = \left(
\begin{array}{ccc}
m&& \\
&0&\\
&&-m
\end{array}
\right),
\end{align}
in the sense of theories on the flat space.

 In the case that $N_{f}=12$, a $\log M$ term appears and we do not have any interpretation of the effective theory. It may be worth emphasizing that a $\log M$ term will appear when $\frac{N_{f}}{3}=2N-2$, where $N_{f}=2N-2$ is the threshold for a ``bad" theory of $\mathcal{N}=4 \ \U(N)$ SQCD with $N_{f}$ flavors.
When $N_f\geq 15$, we find that the massive multiplets simply decouple in the infinite mass limit because we can change the order of the limit of the mass and the integrals. Indeed, 
\begin{align}
\tilde{Z}^{N_f}_{\text{U}(3)}\xrightarrow[m \to \infty]{} \left(\frac{1}{M}\right)^{\frac{3N_f}{3}}\tilde{Z}^{\frac{N_f}{3}}_{\U(3)}\big|_{m=0},
\end{align}
and in this case the IR effective theory corresponds to the theory on the trivial Coulomb branch point
\begin{align}
\sigma = \left(
\begin{array}{ccc}
0&& \\
&0&\\
&&0
\end{array}
\right).
\end{align}

\section{Conclusion and Discussion}
It is known that the IR dynamics of three-dimensional supersymmetric SQCD theories strongly depends on the number of matter multiplets. Hence, it is reasonable that we give infinite mass to matter multiplets in order to decouple them and then investigate the effects. 
We considered the round three-sphere partition function of two types of mass-deformed U($N$) SQCD with massive hypermultiplets and what happens when we take the infinite mass limit. The deformations are following: (i) only massive matter fields, and (ii) massive and massless matter fields.
 Generally speaking, the vacuum (in this paper we only consider the Coulomb branch) must be chosen in order to argue on whether a matter hypermultiplet decouples from a theory from the viewpoint of theories on flat space.
 On the other hand, because the sphere partition function is written in terms of the integrals over all possible Coulomb branch parameters, it seems that we cannot argue on whether the matter hypermultiplets decouple.
  Then, we focused on the large $N$ limit, in order to determine a dominant point of the Coulomb branch. The partition function was evaluated by the solution of the saddle point equation and the solution corresponds to a specific point of the Coulomb branch. 
  Therefore, we could investigate the decoupling of matter fields and the effective IR theory by following the solution. Finally, we confirmed that an effective theory on a non-trivial point of the Coulomb branch appears through the exact calculation of the partition function of finite-rank SQCD.

In case (i), if we consider a theory on the trivial Coulomb branch and take the mass to infinity, then, all massive matter fields decouple from the theory and its partition function diverges. In fact, this argument is not valid because it is not guaranteed that the limit of mass is commutative with the integrals of the matrix model. We found that the solution of the saddle point solution has two separated regions: one is concentrated around $m$ and the other around $-m$. This means that in the infinite mass limit the gauge group U($N$) is broken down to U($\frac{N}{2}$)$\times$U($\frac{N}{2}$) with massless hypermultiplets and FI terms. Even for cases of finite $N$, this picture we obtained from the large $N$ analysis may be true except when $\frac{N}{2}$ is not an integer.

In case (ii), the behavior of the partition function depends on the number of the massless flavors.  When $\frac{N_{f}}{3} > 2N-2$, the limit of the mass is commutable with the integrals and the massive matter fields simply decouple from the theory. This corresponds to the case that the solution of the saddle point equation is concentrated on the origin of the Coulomb branch. In case that $\frac{N_{f}}{3} \leq 2N-2$, we found that the gauge group is broken into three parts and the rank of each gauge group depends on the number of flavors. Through the results in cases of finite $N$, we confirmed that a non-trivial effective theory appears in the infinite mass limit except in a few cases.

Let us comment on more general mass deformations. In this paper, we considered above two mass deformations. It is possible to study more general mass deformations of $\mathcal{N}=4$ U($N$) SQCD as long as mass deformations preserve $\mathcal{N}=4$ supersymmetry because at least, the large $N$ analysis can be applied to the general mass deformations. The result for the more general mass-deformed theories can be inferred from our results. Then, it is expected that the number of the real mass parameters correspond to that of the peaks of the density function and the number of the flavors which have the same real mass parameter determines the rank of the gauge groups which the original gauge group is spontaneously broken into.

We will also provide the model that becomes the ABJM theory in the infinite mass limit with the help of this vacuum selection in the Appendix \ref{ABJM}. This can be regarded as an example that connects a theory whose free energy is proportional to $N^{2}$ to one whose free energy is proportional to $N^{\frac{3}{2}}$ by means of a continuous parameter. This is consistent with the F-theorem  \cite{JKPS}\footnote{
In this argument, we omit the decoupled massive free sectors.
}.

Let us now comment on the F-theorem. In our analysis, it can be verified that the free energies of many theories are divided into two parts in the infinite mass limit as
\begin{align}
\label{decom}
F \rightarrow F_{\text{SCFT}}+F_{\text{Massive}},
\end{align} 
where $F_{\text{SCFT}}$ is the free energy of an interacting SCFT and $F_{\text{Massive}}$ is that representing the sector of free massive multiplets. $F_{\text{Massive}}$ is proportional to $m$ and we can counter it by a local-counter term, which corresponds the Einstein-Hilbert action of $S^{3}$ \cite{CDFKS0}. We expect that
\begin{align}
F_{\text{UV}}>F_{\text{SCFT}},
\end{align}
where $F_{\text{UV}}$ is the free energy when $m=0$ because it can naively be considered that the limit $m\rightarrow \infty$ corresponds to a deep IR limit and $m=0$ corresponds to a UV limit. In fact, this relation is valid at least in our results in Section \ref{finite}. We also remark that we encounter exceptional theories, whose partition functions exhibit $\log M$ behavior in the infinite mass limit, and these theories cannot be interpreted as in \eqref{decom}. This is because the leading behavior of the partition function can be evaluated by substituting the dominant point of the Coulomb branch for the action. Thus, the contributions from the points of the Coulomb branch (at most countable points) cannot cause the logarithmic factors. It may be possible that the logarithmic factors arise from the contributions of the uncountably infinite points of the Coulomb branch. It may be interesting to investigate the IR behavior of such a theory and to consider the F-theorem.

\section*{Acknowledgement}
 K.S would like to thank Masazumi Honda, Tomoki Nosaka and Seiji Terashima for reading of manuscript of this paper and providing us valuable comments and discussion. K.S. also would like to thank Takaya Miyamoto and Shuichi Yokoyama for useful discussions. K.S. is supported by the JSPS fellowship and by Grant-in-Aid for JSPS Fellow No.18J11714.

\appendix

\section{A brief summary of resolvent methods}
\label{res}
Here, we introduce resolvent methods and further details of the calculation of the density function in this paper. We follow the argument of the resolvent in \cite{BIC,Ma,Suyama1}. First, we assume that the eigenvalues become dense in the large $N$ limit and we can take the continuous limit as follows:
\begin{align}
 \frac{i}{N}\rightarrow s\in[0,1],\quad x_{i} \rightarrow x(s),\quad \frac{1}{N}\sum_{i=1}^{N}\rightarrow \int ds. 
\end{align}
We introduce the density function $\rho(x)$ defined as 
\begin{align}
\rho(x) \equiv \frac{ds}{dx},
\end{align}
and impose the normalization condition
\begin{align}
\int_{I}\rho(x)=1,
\end{align}
where $I$ is the interval on which $\rho(x)$ is defined. 
In this study, we consider the following type of the saddle point equation, given as a singular integral equation:
\begin{align}
\alpha \left(\text{P}\int_{I}dy \rho(y)\coth\pi(x-y)\right)=V^{\prime}(x), 
\end{align}
where $\alpha$ is constant. It is convenient for us to take $X=e^{2\pi x}$ and $Y=e^{2\pi y}$ as in this case various techniques are available. The saddle point equation is written as
\begin{align}
\alpha\left(1+\text{P}\int_{\mathcal{C}}\frac{dY}{\pi}\frac{\rho(y)}{X-Y}\right)=V^{\prime}(x),
\end{align}
where $X \in \mathcal{C}=[b,a]$. We introduce an auxiliary function $\omega(Z)$ as
\begin{align}
\omega(X)\equiv \alpha\left(1+\int_{\mathcal{C}}\frac{dY}{\pi}\frac{\rho(y)}{X-Y}\right).
\end{align}
This function is defined on all of the complex plane except on $\mathcal{C}$, where $\omega(X)$ has a discontinuity when we across the interval $\mathcal{C}$. The function satisfies the following properties:
\begin{align}
\label{analy1}
&\lim_{X\rightarrow 0}\omega(X)=-\alpha ,\quad \lim_{X\rightarrow \infty} \omega(X)=\alpha, \\
\label{analy2}
\rho(x)=&-\frac{1}{2\alpha i}\lim_{\epsilon \rightarrow 0}\left(\omega(X+i\epsilon)-\omega(X-i\epsilon)\right),\quad X\in \mathcal{C}, \\
\label{analy3}
V^{\prime}(x)=&\frac{1}{2}\lim_{\epsilon \rightarrow 0}\left(\omega(X+i\epsilon)+\omega(X-i\epsilon)\right),\quad X\in \mathcal{C}.
\end{align}
Here, we give a proof of \eqref{analy2} and \eqref{analy3}, which relies on from the discontinuity of $\omega(X)$. The following relation is obtained by changing the integral contour: 
\begin{align}
\int_{\mathcal{C}}\frac{dY}{\pi}\frac{\rho(y)}{X+i\epsilon-Y}=\left(\int_{b}^{X-\epsilon}+\int_{X+\epsilon}^{a}\right)\frac{dY}{\pi}\frac{\rho(y)}{X-Y}+\int_{C^{-}_{\epsilon}}\frac{dY}{\pi}\frac{\rho(y)}{X-Y},\quad (X \in \mathcal{C}),
\end{align}
where $C_{\epsilon}^{-}$ is a circle with radius $\epsilon$ around $Y=X$ in the lower half plane, which is oriented counterclockwise. By the definition of the principal value integral and the residue theorem, we finally obtain  
\begin{align}
\lim_{\epsilon \rightarrow 0}\omega(X+i\epsilon)=\alpha\left(1+\text{P}\int_{\mathcal{C}}\frac{dY}{\pi}\frac{\rho(y)}{X-Y}-i\rho(x)\right).
\end{align}
By the same calculation, we also obtain
\begin{align}
\lim_{\epsilon \rightarrow 0}\omega(X-i\epsilon)=\alpha\left(1+\text{P}\int_{\mathcal{C}}\frac{dY}{\pi}\frac{\rho(y)}{X-Y}+i\rho(x)\right).
\end{align}
Thus, the equations \eqref{analy2} and \eqref{analy3} are proved.

From the analyticity\footnote{
We should consider the resolvent $\omega(X)$ such that its branch cut is on $[b,a]$ and it satisfies the asymptotic equations (A.7). Then, we should take the resolvent $\omega(X)$ that has the product of the square root $\sqrt{X-a}\sqrt{X-b}$, not $\sqrt{(X-a)(X-b)}$ because indeed, $\sqrt{(X-a)(X-b)}$ has the branch cut on $[b,a]$, but does not satisfy the asymptotic behavior at $X\to 0$
}, the resolvent is given by 
\begin{align}
\omega(X)=\oint_{C}\frac{dZ}{2\pi i}\frac{V^{\prime}(z)}{X-Z}\frac{\sqrt{(X-a)}\sqrt{(X-b)}}{\sqrt{(Z-a)}\sqrt{(Z-b)}}, \quad Z=e^{2\pi z},
\end{align}
where $C$ is a circle which encloses $\mathcal{C}$. The density function is determined once the potential $V^{\prime}(z)$ is given. We assume that the $n_{i}$ degree poles $X_{0i},(i=1,\dots n_{0})$ of $V^{\prime}(x)$, exist outside of $\mathcal{C}$. We deform the integral contour $\mathcal{C}$ to infinity and pick up the poles $Z=X$ and $Z=X_{0i},\ (i=1,\dots n_{0})$. Thus, the resolvent is written as
\begin{align}
\label{omega1}
\omega(X)&=-\oint_{\infty}\frac{dZ}{2\pi i}\frac{V^{\prime}(z)}{X-Z}\frac{\sqrt{(X-a)}\sqrt{(X-b)}}{\sqrt{(Z-a)}\sqrt{(Z-b)}} \nonumber \\
&=V^{\prime}(x)-\sum_{i=1}^{n_{0}}\text{Res}\left(\frac{V^{\prime}(z)}{X-Z}\frac{\sqrt{(X-a)}\sqrt{(X-b)}}{\sqrt{(Z-a)}\sqrt{(Z-b)}},X_{0i}\right).
\end{align}
To determine the edge of the cut $\mathcal{C}$, we simply solve the equation \eqref{analy1} with the resolvent $\omega(X)$ obtained in \eqref{omega1}.

\section{Mixed Chern-Simons terms}
\label{mix}
It is known that the various Chern-Simons terms exist in three dimensions。 These not only consist of dynamical gauge fields, but also background fields that couple with the current of the global symmetry. These Chern-Simons terms must appear in the infinite mass limit as one-loop effects by integrating out the massive fermions charged under the corresponding symmetries. In particular, on $S^{3}$ we can consider background vector fields that couples with the R-symmetry current and Chern-Simons terms including the background fields. These are important in order to understand what remains after taking the infinite mass limit \cite{CDFKS,CDFKS0,CDFK}. The Chern-Simons terms that will appear are flavor-R and gauge-R mixed Chern-Simons terms given by
\begin{align}
\label{FR}
S^{\text{FR}}_{\text{CS}} \sim &\frac{k_{\text{FR}}}{2\pi}\int_{S^3}\sqrt{g}d^3x\left(\sigma_{\text{f}} +iD_{\text{f}}\right), \\
\label{GR}
S^{\text{GR}}_{\text{CS}}\sim &\frac{k_{GR}}{2\pi}\text{Tr}\int_{S^3} \sqrt{g}d^3x\left(\sigma+iD\right),
\end{align}
where $\sigma_{\text{f}}$ and $D_{\text{f}}$ represent the scalar and auxiliary fields of the background vector superfields, respectively. Here, we only write the parts of the action that contribute to it after applying localization methods. The induced Chern-Simons levels are given by integrating out a Majorana fermion $\psi$ as
\begin{align}
k^{\text{FR}}_{\psi}=\frac{\Delta_{\psi}}{2}\sgn(M_{\psi})\sum_{\text{f}}q_{\psi,\text{f}},\\
k^{\text{GR}}_{\psi}=\frac{\Delta_{\psi}}{2}\sgn(M_{\psi})\sum_{i}q_{\psi,i},
\end{align}
where $q_{\text{f}}$, $q_{\text{g}}$ and $\Delta$ correspond to flavor, gauge and R charges respectively. Furthermore, $M_{\psi}$ is the effective real mass of the fermions on a point of the Coulomb branch. In this paper, $M_{\psi}=\sum_{\text{f}}q_{\text{f}} \sigma_{\text{f}}+\sum_{i}q_{i}\sigma_{i}$, where $i$ labels the U(1) gauge groups on the Coulomb branch.

We use the terms \eqref{FR} and \eqref{GR} after applying a localization technique. These are given by
\begin{align}
e^{S^{\text{FR}}_{\text{CS}}}=e^{2\pi k_{\text{FR}}\sigma_{f}},\\
e^{S^{\text{GR}}_{\text{CS}}}=e^{2\pi k_{\text{GR}}\sigma},
\end{align}
where the supersymmetric configuration of the background fields and the localization locus are required:
\begin{align}
D_{f}=-i\sigma_{f}, \quad D=-i\sigma,\quad \text{(Other fields)}=0.
\end{align} 
The real mass is given by the expectation value of the background field $\sigma_{f}=m$. Thus, the flavor-R Chern-Simons terms \eqref{FR} induced in the infinite mass limit can corresponds to the contributions from the free massive degrees of freedom. The induced gauge-R Chern-Simons term \eqref{GR}
corresponds to the FI term and the contributions from massive degrees of freedom when we shift $\sigma$ by $m$.

As an example, we attempt to obtain the decoupled free massive sector and the FI terms of the U(2) SQCD case described in Section \ref{u2sqcd} from the induced Chern-Simons terms\footnote{We would like to thank Masazumi Honda for giving us useful comments and discussion on this point.}. We assume that the classical Coulomb branch parameters $(\sigma_1,\sigma_2)$ are written as $(-m-\delta\sigma_1,m-\delta\sigma_2)$. Then, the gauge group U(2) is broken down to U(1$)_{L}$$\times$ U(1$)_{R}$. The effective mass and the charges of the massive gauginos and Majorana fermions of chiral multiplets are summarized in  Table \ref{charge} \footnote{
An $\mathcal{N}=4$ gauge theory has a chiral multiplet in the adjoint representation of the gauge group. Therefore, it seems that we must consider the contributions from the chiral multiplets. However, the canonical R-charge of the chiral multiplet is 1. Thus, the R-charge of the fermion component is 0 and it does not contribute to the gauge-R mixed Chern-Simons level. 
}.

\begin{table}[ht!]
\begin{center}
\begin{tabular}{|c|c|c|c|}\hline
 & effective mass & U(1$)_R$ & U(1$)_{L}$ $\times$ U(1$)_{R}$ \\ \hline
$\lambda_{+}$ & $\sigma_{1}-\sigma_{2}$ & 1& $(1,-1)$ \\
$\lambda_{-}$ & $\sigma_{2}-\sigma_{1}$ & 1& $(-1,1)$ \\
$\psi_{1\pm} $  &  $\pm m+\sigma_{1}$& $ -\frac{1}{2}$ & $(1,0)$ \\
$\psi_{2\pm} $  & $ \pm m+\sigma_{2} $& $-\frac{1}{2} $ & $(0,1)$ \\
$\widetilde{\psi}_{1\pm}$ & $\pm m -\sigma_{1} $ & $-\frac{1}{2}$ & $(-1,0)$\\
$\widetilde{\psi}_{2\pm}$ & $\pm m -\sigma_{2} $ & $-\frac{1}{2}$&  $(0,-1)$ \\ \hline
\end{tabular}
\caption{The effective mass, R-charge and gauge charge of the fermions under U(1)$\times$U(1). Here, $\lambda_{\pm}$ denote gauginos and $\psi$ ($\widetilde{\psi}$) is a Majorana fermion in the chiral multiplet in the fundamental (anti-fundamental) representation.}
\label{charge}
\end{center}
\end{table}

 The contributions of the massive gauginos to the induced Chern-Simons terms as follows:
\begin{align}
\lambda_{+}&:\ e^{\pi \Delta_\lambda\text{sign}\left(-2m-\delta \sigma_1+\delta \sigma_2 \right)\left(-m-\delta\sigma_1-(-\delta\sigma_2+m)\right)}\\
\lambda_{-}&:\ e^{\pi \Delta_\lambda\text{sign}\left(2m+\delta\sigma_1-\delta \sigma_2 \right)\left(-(-m-\delta\sigma_1)+(-\delta\sigma_2+m)\right)}.
\end{align}
Thus, the total contributions of massive gauginos when $m\to \infty$ are 
\begin{align}
e^{2\pi\Delta_\lambda(2m+\delta\sigma_1-\delta\sigma_2 )}.
\end{align} 
The first term can be interpreted as the massive free part and the second and third terms are the induced FI terms of U(1)$_{L}$ and U(1$)_{R}$.

The contributions of the massive matter fermions are summarized as follows:
\begin{align}
\psi_{1-}:\ &e^{\pi\Delta_\psi\sgn(-2m-\delta\sigma_1)\left(-2m-\delta\sigma_1\right)}\\
\psi_{2+}:\ &e^{\pi\Delta_\psi\sgn(2m-\delta\sigma_2)\left(2m-\delta\sigma_2\right)}\\
\widetilde{\psi}_{1+}:\ &e^{\pi\Delta_\psi\sgn(2m+\delta\sigma_1)(2m+\delta\sigma_1)}\\
\widetilde{\psi}_{2-}:\ &e^{\pi \Delta_\psi\sgn(-2m+\delta\sigma_2)(-2m+\delta\sigma_2)}.
\end{align} 
Then, the total contribution of the massive matter fermions to the induced Chern-Simons term is given by 
\begin{align}
e^{\pi\frac{N_f}{2}\Delta_\psi(8m+2\delta\sigma_1-2\delta\sigma_2)}.
\end{align}
The first term can be interpreted as representing the contributions from the massive free sector and the second and third can be interpreted as FI terms. Then, we conclude that in this case the total contributions from the free massive sectors when $m\to \infty$ is given by 
\begin{align}
e^{2m\pi\left(2-N_f\right)},
\end{align}
and the total induced FI terms are given by
\begin{align}
&e^{\pi\left(2-\frac{N_f}{2}\right)\delta\sigma_1}, \quad \text{for U(1)$_{L}$}, \\
&e^{-\pi\left(2-\frac{N_f}{2}\right)\delta\sigma_2}, \quad \text{for U(1)$_{R}$}.
\end{align}
These results are same as those we obtained in Section \ref{u2sqcd} from the calculation of the matrix model. Thus, the effects that appear in the infinite mass limit in the matrix model can indeed be regarded as the induced mixed Chern-Simons terms. The result can easily be generalized to other theories in this paper. These consequences are not surprising because the one-loop parts of the vector and chiral multiplets in the matrix model must inherit such one-loop effects after integrating out massive fermions.

\section{Convergence bound of matrix models}
\label{conv}
In this section, we discuss the convergence of the matrix models. The convergence bound of the matrix model of SQCD was first discussed in \cite{WY}. In \cite{SKL}, it is also pointed out that the convergence bound is indistinguishable from the unitarity bound of the monopole operator in the Veneziano limit.

 We consider the convergence of the matrix model introduced in \eqref{FImat}: 
\begin{align}
Z=\frac{1}{N!}\int \prod_{i=1}^{N}dx_{i}\frac{e^{\pi \zeta \sum_{i}x_{i}}\prod_{i<j}4\sinh^2\left(\pi (x_{i}-x_{j})\right)}{\prod_{i}\bigg(2\cosh\pi (x_{i})\bigg)^{N_{f}}}.
\end{align}
In order to check whether the integral is convergent, it is suffiient to know the asymptotic behavior of the integrand when we take one of the integral valuables $|x_{i}|\rightarrow \infty$. Thus, we focus on $x_{1}$ and study the asymptotic behavior of the integrand. When $|x_{1}|\rightarrow \infty$, the part of  the integrand that is related to the convergence is
\begin{align}
e^{\pi |x_{1}|\left(\text{sign}(x_{1})\zeta+2(N-1)-N_{f}\right)}.
\end{align} 
Thus, for the matrix model to converge the relation 
\begin{align}
\label{bound}
|\zeta|+2(N-1)-N_{f}<0
\end{align}
must hold. This threshold corresponds to the condition that the solution of the saddle point equation in \eqref{FImat} in the large $N$ limit exists. We note that the matrix models of the effective theories \eqref{effectivematrix} and \eqref{effectivematrix2} satisfy the above relation and converge. In fact, each matrix model narrowly satisfies the bound. For example, for the case of \eqref{effectivematrix}, the left-hand side of the bound \eqref{bound} is $-2$, which does not depends on any parameter. Therefore, the convergence of the matrix model restricts the theory that appears in the infinite mass limit. In fact, in the subsection \ref{32} we assume that the solution of the saddle point equation where the gauge group U($N$) is broken down to U($N_{1}$)$\times$ U($N_{2}$) $(N_{1}>N_{2})$\footnote{
We assume this situation without loss of generality.
} is allowed. Then, the condition for the convergence of the matrix model corresponding to the effective theory is given by 
\begin{align}
0>&\frac{N_{f}}{2}-2N_{2}+2(N_{1}-1)-\frac{N_{f}}{2}=2(N_{1}-N_{2})-2, \\
0>&\bigg| \frac{N_{f}}{2}-2N_{1}\bigg|+2(N_{1}-1)-\frac{N_{f}}{2}.
\end{align}
The first line is not satisfied when $N_{1}>N_{2}$. Therefore, only the case that $N_{1}=N_{2}$ is allowed owing to the convergence of the matrix model of the effective theory\footnote{
We would like to thank Tomoki Nosaka for pointing out and discussing this point.}. By the same argument concerning the convergence bound, we can also understand why $N_{1},\ N_{2}$ and $N_{3}$ satisfy the relation \eqref{rank}.

\section{ABJM theory as an effective theory}

\label{ABJM}

Here, we consider the theory whose effective theory in the large mass limit is the ABJM theory. Naively speaking, the theory when $m=0$ corresponds to the UV theory in the sense that the energy scale that is determined by the radius of the three-sphere is significantly bigger than the mass parameter. In the same sense, the theory in the infinite mass limit corresponds to the IR theory in the same sense. The SYM theory we introduce here flows to the ABJM theory in the above sense. Which effective theories appear depends on the mass assignment to matter multiplets and the representation of the matter fields. It is possible to anticipate that the ABJM theory will appear as the effective theory in the infinite mass limit using the insight developed so far.

We consider the U($2N$) SYM theory with two massive hypermultiplets in the adjoint representation and $2N_{f}$ massive fundamental hypermultiplets. The matrix model is given by 
\begin{align}
\label{orimatrix}
Z=\frac{1}{(2N)!}\int{d^{N}x}&\prod_{i>j}^{2N}\frac{4\sinh^{2}\pi(x_{i}-x_{j})}{\left(2\cosh\pi(x_{i}-x_{j}+2m)2\cosh\pi\left(x_{i}-x_{j}-2m\right)\right)^{2}}\nonumber \\
\times&\prod^{2N}_{i}\frac{1}{\left(2\cosh\pi\left(x_{i}+m\right)2\cosh\pi\left(x_{i}-m\right)\right)^{N_{f}}},
\end{align}
where it is necessary to give the adjoint hypermultiplets real mass $\pm 2m$ and the hypermultiplets real mass $\pm m$. Then, we assume that the saddle point configuration is splitting, which means that the saddle point solution ${x_{0i}}$ has the following separated region:
\begin{align}
\begin{cases}
x_{0i}=m+\lambda_{i}\quad i\in 1,\dots, N,\\
x_{0i}=-m+\tilde{\lambda}_{i} \quad i\in 1,\dots, N
\end{cases}.
\end{align}
Under this assumption, the free energy $F=-\log Z$ is evaluated for the solution $x_{0i}$ as
\begin{align}
F=&-\sum_{i>j}\bigg[\log4\sinh^2\pi\left(\lambda_{i}-\lambda_{j}\right)+\log4\sinh^2\pi\left(\tilde{\lambda}_{i}-\tilde{\lambda}_{j}\right)\bigg]+2\sum_{i,j}\log2\cosh\pi\left(\tilde{\lambda}_{i}-\lambda_{j}\right)\nonumber \\ 
&+N_{f}\sum_{i}\log2\cosh\pi\lambda_{i}+N_{f}\sum_{i}\log2\cosh\pi\tilde{\lambda}_{i}  \nonumber \\
&-\sum_{i>j}\bigg[\log4\sinh^2\pi\left(\tilde{\lambda}_{i}-\lambda_{j}+2m\right)-2\log\cosh\pi\left(\tilde{\lambda}_{i}-\lambda_{j}+4m\right)\bigg]\nonumber \\
&+2\sum_{i>j}\log2\cosh\pi\left(\lambda_{i}-\lambda_{j}+2m\right)2\cosh\pi(\lambda_{i}-\lambda_{j}-2m)\nonumber \\
&+2\sum_{i>j}\log2\cosh\pi\left(\tilde{\lambda}_{i}-\tilde{\lambda}_{j}+2m\right)2\cosh\pi(\tilde{\lambda}_{i}-\tilde{\lambda}_{j}-2m)\nonumber \\
&+N_{f}\sum_{i}\log2\cosh\pi\left(\lambda_{i}+2m\right)+N_{f}\sum_{i}\log2\cosh\pi\left(\tilde{\lambda}_{i}-2m\right).\nonumber \\
=& -\sum_{i>j}\bigg[\log4\sinh^2\pi\left(\lambda_{i}-\lambda_{j}\right)+\log4\sinh^2\pi\left(\tilde{\lambda}_{i}-\tilde{\lambda}_{j}\right)\bigg]+2\sum_{i,j}\log2\cosh\pi\left(\tilde{\lambda}_{i}-\lambda_{j}\right) \nonumber \\
&+N_{f}\sum_{i}\log2\cosh\pi\lambda_{i}+N_{f}\sum_{i}\log2\cosh\pi\tilde{\lambda}_{i}+\text{(massive part)}.
\end{align}
The massive part in the infinite mass limit is given by $N_{0}m$ where $N_{0}$ is some constant. The massless part of the action corresponds to the free energy of the U($N$) $\times$ U($N$) quiver SYM with two bi-fundamental multiplets and $N_{f}$ fundamental hypermultiplets multiplets charged under each U($N$), which is evaluated using saddle point approximation. The matrix model of this effective theory is given by
\begin{align}
\label{effABJM}
Z_{\text{eff}} \sim \int d^{N}\lambda d^{N}\tilde{\lambda}\displaystyle\frac{\prod_{i>j}4\sinh^2\pi(\lambda_{i}-\lambda_{j})\sinh^2\pi(\tilde{\lambda}_{i}-\tilde{\lambda}_{j})}{\prod_{i,j}\left(2\cosh\pi\left(\tilde{\lambda}_{i}-\lambda_{j}\right)\right)^2\prod_{i}\left(2\cosh\pi\lambda_{i}2\cosh\pi\tilde{\lambda}_{i}\right)^{N_{f}}}.
\end{align}
 The massive part is proportional to $N^{2}m$ when we consider that the mass $m$ is considerably bigger than the typical order of the eigenvalues. We will show that this matrix model is the same as the square of the matrix model of the ABJM theory with the Chern-Simons level $k=N_{f}$ in the large $N$ limit. We will solve the saddle point equation of \eqref{effABJM} by following the approach in \cite{HKPT}, where the eigenvalues are proportional to $\sqrt{N}$ in the large $N$ limit with the Chern-Simons level $k$ kept finite.

We assume that the saddle point configuration satisfies the condition 
\begin{align}
\label{ansatz1}
\lambda_{i}=\tilde{\lambda}_{i}.
\end{align}
This is plausible in the sense that the action $S(\lambda,\tilde{\lambda})$ is invariant under exchange of $\lambda$ and $\tilde{\lambda}$. Under this assumption, it is sufficiently to consider the saddle point equation of the matrix model 
\begin{align}
\tilde{Z}_{\text{eff}}=&\frac{1}{(2N)!}\int d^{N}\lambda\displaystyle\frac{\prod_{i>j}4\sinh^2\pi(\lambda_{i}-\lambda_{j})}{\prod_{i,j}\left(2\cosh\pi\left(\lambda_{i}-\lambda_{j}\right)\right)\prod_{i}\left(2\cosh\pi\lambda_{i}\right)^{N_{f}}}\\
\equiv & \int d^{N}\lambda e^{-\tilde{S}_{\text{eff}}(\lambda)}. \nonumber
\end{align}
For $N_{f}=1$, this matrix model is known as the mirror dual matrix model of the ABJM theory with $k=1$ \cite{KWY2}. This matrix model is studied in \cite{MP,GM} in the Veneziano limit where $N_{f}$ is taken to infinity while $\frac{N_{f}}{N}$ is kept finite. In this paper, we do not employ the Veneziano limit. Rather, we take $N$ to infinity while keeping $N_{f}$ finite.

We evaluate the action, which is explicitly written as 
\begin{align}
\label{action}
\widetilde{S}_{\text{eff}}(\lambda)=-\sum_{i>j}\log4\sinh^2\pi\left(\lambda_{i}-\lambda_{j}\right)+\sum_{i,j}\log2\cosh\pi\left(\lambda_{i}-\lambda_{j}\right)+N_{f}\sum_{i}\log2\cosh\pi\left(\lambda_{i}\right).
\end{align}
Here, we take the continuous limit in the large $N$ limit. We define the continuous parameter $s$ resulting from the label of the eigenvalues as $s=\frac{i}{N}+s_{b}$. The continuous value $s$ runs from $s_{b}$ to $s_{b}+1$, where $s_{b}$ is constant. The eigenvalues are replaced by a function of $s$ which is monotonically increasing and differentiable. The summation is replaced by an integral as
\begin{align}
\sum_{i}\rightarrow N\int_{s_{b}}^{s_{b}+1}ds,
\end{align} 
where we do not introduce the density function. We assume that an ABJM-type ansatz for the eigenvalues which are proportional to $\sqrt{N}$ in the large $N$ limit as \cite{HKPT}
\begin{align}
\lambda(s)=\sqrt{N}x(s),
\end{align}
From the above expression, the final term of the action \eqref{action} is evaluated as
\begin{align}
N_{f}\sum_{i}\log2\cosh\pi\left(\sqrt{N}x_{i}\right)\rightarrow N^{\frac{3}{2}}N_{f}\pi \int_{s_{b}}^{s_{b}+1}|x(s)|.
\end{align}
The evaluation of the first and second terms of the action is non-trivial as the naive order of these part is $N^{2}$, which reduces to $N^{\frac{3}{2}}$. We briefly review this using the technique  developed in \cite{NST2}. We rewrite the first term as follows:
\begin{align}
\label{act1}
&\sum_{i>j}\log4\sinh^2\pi\left(\lambda_{i}-\lambda_{j}\right)\nonumber \\
\rightarrow& N^{2}\int\int_{s>s'} dsds^{\prime}\log4\sinh^2\pi\left(\sqrt{N}(x-x')\right)\nonumber \\
=&\frac{N^{2}}{2}\int\int dsds'\bigg[2\pi\sqrt{N}|x-x'|+\log\left[4\sinh^2\pi\left(\sqrt{N}(x-x')\right)e^{-2\pi\sqrt{N}|x-x'|}\right]\bigg],
\end{align}
\begin{align}
\label{act2}
&\sum_{i,j}\log2\cosh\pi\left(\lambda_{i}-\lambda_{j}\right)\nonumber  \\
\rightarrow &N^{2}\int\int dsds^{\prime}\log2\cosh\pi\left(\sqrt{N}(x-x')\right) \nonumber \\
=&N^{2}\int\int dsds'\bigg[\pi\sqrt{N}|x-x'|+\log\left[2\cosh\pi\left(\sqrt{N}(x-x')\right)e^{-\pi\sqrt{N}|x-x'|}\right]\bigg],
\end{align}
where $x$ and $x'$ denote $x(s)$ and $x(s')$, respectively. The first terms in \eqref{act1} and \eqref{act2} cancel. The second terms in \eqref{act1} and \eqref{act2} are evaluated by the following approximation formulae:
\begin{align}
 \int_{s_0} ds \log\left(1\pm e^{-2z(s)}\right) &\sim \frac{1}{\sqrt{N} \dot{x}(s_{0}) }\int_{C_+} dt \log\left(1\pm e^{-2t}\right), \label{logcoshapprox1} \\
 \int^{s_0} ds \log\left(1\pm e^{+2z(s)}\right) &\sim\frac{1}{\sqrt{N} \dot{x}(s_{0}) }\int_{C_-}dt \log\left(1\pm e^{-2t}\right),
\label{logcoshapprox2}
\end{align}
for
$\dot{x}(s)|_{s=s_0}>0$
where
\begin{align}
 z(s)=\sqrt{N} x(s)+v(s),
\end{align}
$x(s_0)=0$ and the path $C_{\pm}$ is a straight line between 
$t=\pm v(s_0)$ and $t=\sqrt{N} \dot{x}(s_{0})$ with $N \rightarrow \infty$. In our case, $v(s)=0$ and $u(s)$ is real. The contour $C_{\pm}$ is a half line from 0 to $\infty$. Thus, the remaining parts of the free energy are
\begin{align}
& N^{\frac{3}{2}} \int 
\frac{ds^{\prime}}{\pi \dot{x}(s')} \left(
-2 \int_0^{\infty} dt \log \left(\sinh(t)e^{-t}\right)) + \int_{0}^{\infty} dt \log \left(\cosh(t)e^{-t}\right)+ \int_{0}^{\infty} dt \log \left(\cosh(t)e^{-t}\right)\right) \nonumber \\
=& N^{\frac{3}{2}} \int \frac{ds^{\prime}}{\pi \dot{x}(s')} \left(-2 \int_0^{\infty} dt \log (\frac{\sinh (t)}{\cosh (t)} )\right) \nonumber \\ 
=& N^{\frac{3}{2}} \int \frac{ds^{\prime}}{\dot{x}(s')}\frac{1}{4} \pi,
\end{align}
where we have assumed that $\dot{x}_1(s')>0$ and that there are no singularities in the $t$-plane when deforming the contour $C_\pm$. However, there are singularities in the action where a $\cosh$ factor vanishes. We can observe that if 
\begin{align}
 -\frac{1}{4} < {\rm Im} (v(s))- {\rm Re} (v(s)) 
\frac{{\rm Im}(\dot{x}(s)) }{{\rm Re} (\dot{x}(s))} < \frac{1}{4},\quad s \in [s_{b},s_{b}+1]
\label{cz}
\end{align}
then there is no obstruction to the deformation of the contour.
If this is not the case, 
then, we can shift $z_2 \rightarrow z_2+ i n/2$, where $n$ is an integer, in order to satisfy the condition (\ref{cz}). 
In this case, the above bound \eqref{cz} is always satisfied since $\text{Im}\left({x(s)}\right)=0$. Plugging the above expressions into the action \eqref{action}, we obtain the leading part of the free energy as 
\begin{align}
\label{freefinal}
F[x]=\pi N^{\frac{3}{2}}\int_{s_b}^{s_{b}+1}ds\left[N_{f}|x(s)|+\frac{1}{4\dot{x}(s)}\right].
\end{align} 
The saddle point equation is given by
\begin{align}\
\label{sadABJM}
0=\frac{\delta F[x] }{\delta x(s) }=N_{f}\text{sign}(x(s))+\frac{1}{4}\frac{d}{ds}\frac{1}{(\dot{x}(s))^2},
\end{align}
with the boundary condition
\begin{align}
\frac{1}{(\dot{x}(s))^2}\bigg|_{\text{boudary}}=0.
\end{align}
The solution of the equation \eqref{sadABJM} is potentially discontinuous at the zeros of $x(s)$ because there a sign function in \eqref{sadABJM}. However, we must make $x(s)$ to be continuous everywhere in $(s_b,s_{b}+1)$ as we assumed that $x(s)$ is differentiable. We must take this fact into account and the boundary conditions are explicitly written as
\begin{align}
\label{bdc1}
\frac{1}{\dot{x}(s)}\bigg|_{s=s_{b}}=&0,\\
\label{bdc2}
\frac{1}{\dot{x}(s)}\bigg|_{s=s_{b}+1}=&0
\end{align} 
where these equations arise from the edge of the domain of $x(s)$. Then, the solution of the saddle point equation \eqref{sadABJM} is\footnote{
There is choice of the over all sign $\pm$ due to the square root. We choose the sign so that the solution monotonically increases.
}
\begin{align}
\label{solABJM}
x(s)&=\frac{\text{sign}\left(s-s_{b}-\frac{1}{2}\right)}{\sqrt{2N_{f}}}\left[1-\sqrt{1-2\bigg|s-s_{b}-\frac{1}{2}\bigg|}\right],\\
&\frac{ds}{dx}\equiv\rho(x)=2N_f\left(\frac{1}{\sqrt{2N_f}}-|x|\right).
\end{align}
It is convenient to take $\tilde{s}\equiv s-s_{b}-\frac{1}{2}$ and we rewrite $\tilde{s}\ \in [-\frac{1}{2},\frac{1}{2}]$ as $s$. Consequently, we can evaluate the free energy by plugging this solution into \eqref{freefinal} as 
\begin{align}
F=\pi N^{\frac{3}{2}}\int_{-\frac{1}{2}}^{\frac{1}{2}} ds\left[\sqrt{\frac{N_{f}}{2}}\left(1-\sqrt{1-2|s|}\right)+\sqrt{\frac{N_{f}}{8}}\sqrt{1-2|s|}\right]=\frac{\pi\sqrt{2N_{f}}N^{\frac{3}{2}}}{3}.
\end{align}

\begin{figure}[h!]
\begin{minipage}{0.5\hsize}
 \begin{center}
 \includegraphics[width=8cm]{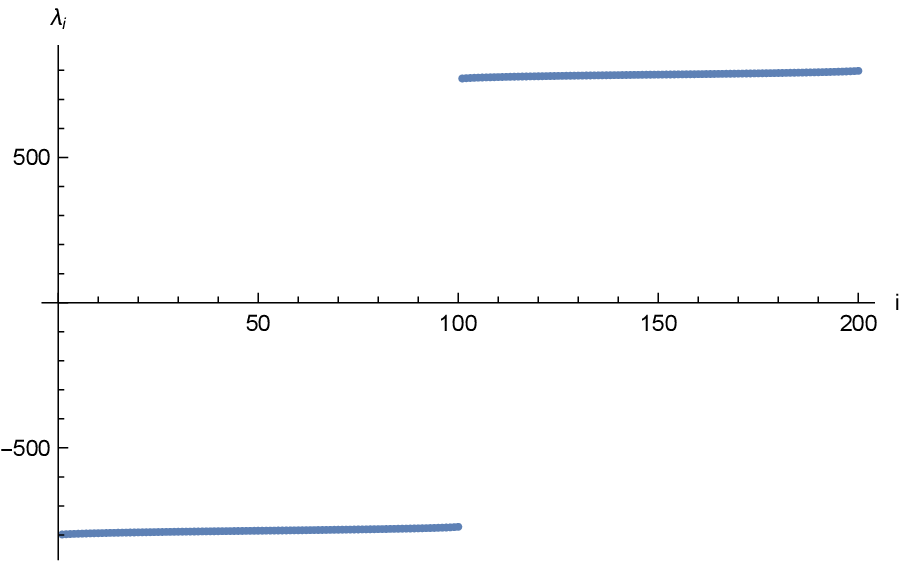}
 \end{center}
 \end{minipage}
 \begin{minipage}{0.5\hsize}
 \begin{center}
 \includegraphics[width=8cm]{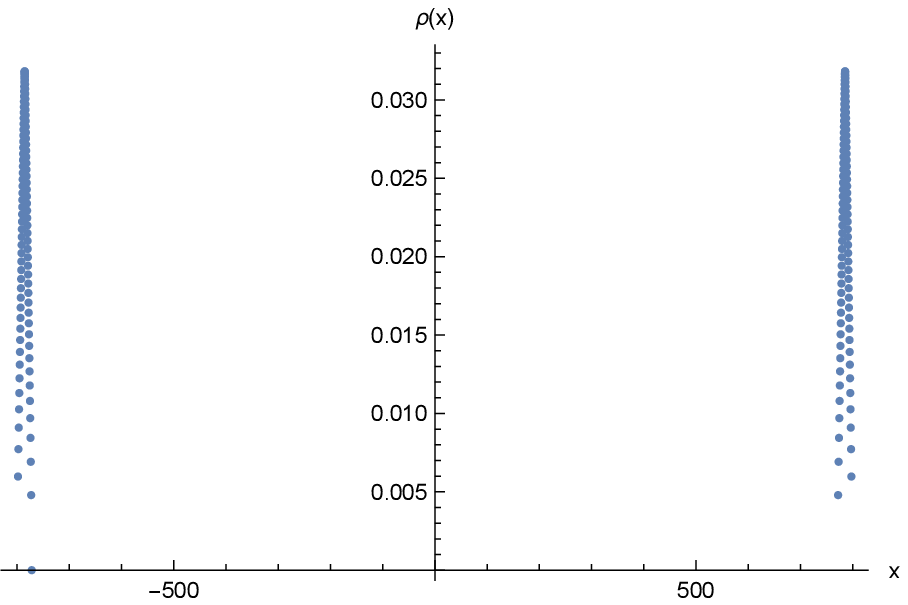}
 \end{center}
 \end{minipage}
 \caption{The left figure shows the solution of the saddle point equation of \eqref{orimatrix} with $(N,N_f,m)=(200,2,500)$. The horizontal line means the label of the eigenvalues. The right one shows that the density of the eigenvalues with the same parameter. The horizontal line means the degree of the eigenvalues.}
 \label{fig1}
 \end{figure} 
 
\begin{figure}[h!]
\begin{minipage}{0.5\hsize}
 \begin{center}
 \includegraphics[width=8cm]{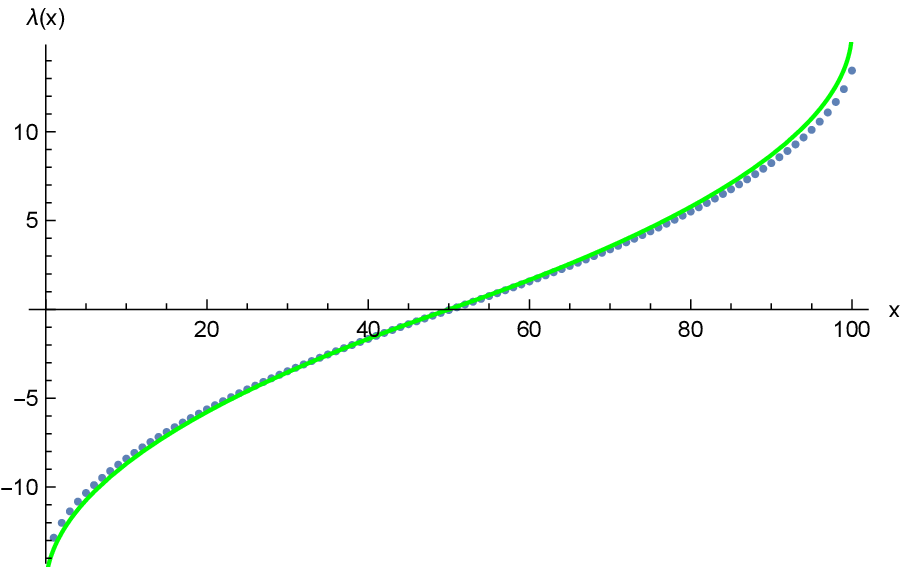}
 \end{center}
 \end{minipage}
 \begin{minipage}{0.5\hsize}
 \begin{center}
 \includegraphics[width=8cm]{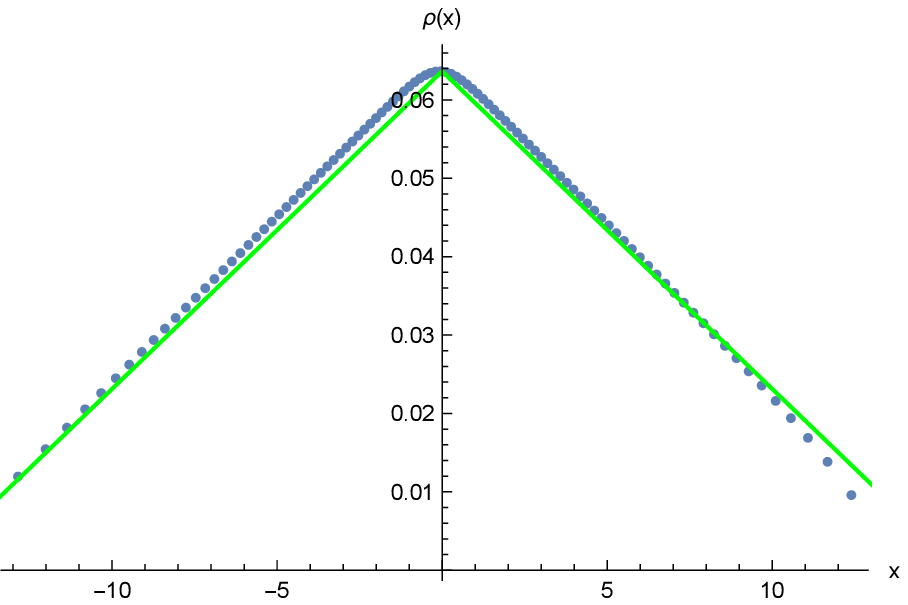}
 \end{center}
 \end{minipage}
 \caption{The left figure shows the comparison of the numerical solution (Blue dots) of the saddle point equation of \eqref{orimatrix} with the analytic solution \eqref{solABJM} (Green line). The right one shows that the comparison of the density function from numerical analysis(Blue dots) with the analytic one (Green line). In this figures we focus on the eigenvalues of the first $\frac{N}{2}$ eigenvalues and shifted the valuable $x$ by $m$.}
 \label{fig2}
 \end{figure}

We can see that the number of the fundamental flavors $N_{f}$ of the SYM theory corresponds to the Chern-Simons level of the ABJM theory at least in the large $N$ limit as the free energy of the ABJM theory with Chern-Simons level $k$ is $\frac{\pi\sqrt{2k}}{3}N^{\frac{3}{2}}$. We present the numerical solution of the saddle point equation of \eqref{orimatrix} in the large $N$ and large mass region and compare it with the analytic solution \eqref{solABJM}. From the following figure, we can observe that the solution has two separated regions and in fact the distance between the two regions is given by $m$. Therefore, we conclude that in the large $N$ and large mass limit with maintained $\sqrt{N}\ll m$, the free energy of \eqref{orimatrix}  
\begin{align}
F=2F_{\text{ABJM}}(N,k=N_{f})+F_{\text{massive}},
\end{align}
where $F_{\text{ABJM}}(N,k)$ is the free energy of the U($N)_{k}$ $\times $U($N)_{-k}$ ABJM theory and $F_{\text{massive}}$ is the free energy of the free massive sector, which originates from massive hyper and vector multiplets. Here, $F_{\text{massive}}$ is proportional to $N^{2}$ while $F_{\text{ABJM}}$ is proportional to $N^{\frac{3}{2}}$.
The free energy of the IR effective theory is $2F_{\text{ABJM}}$. This is consistent with the F-theorem 
\footnote{
In the infinite mass limit, $F_{\text{Massive}}$ is proportional to mass and this term is a scheme dependent term in the three-dimensional theory. Thus we can counter this term by a local counter term $\Lambda\int_{S^3}\sqrt{g}(R+\cdots)$, where $\Lambda$ has mass dimension one. 
} in the sense that the free energy of the UV theory, which corresponds to taking $m=0$, is proportional to $N^{2}$ while that of the deep IR theory, which corresponds to taking $m=\infty$, is proportional to $N^{\frac{3}{2}}$. Thus, we conclude that this model \eqref{orimatrix} is an example that connects a theory whose free energy is proportional to  $N^{2}$ to one whose free energy is proportional to $N^{\frac{3}{2}}$ through a continuous parameter.

\end{document}